\documentclass[final,5p,times,twocolumn]{elsarticle}
\pdfoutput=1

\usepackage{graphicx}  
\usepackage{url}
\usepackage{caption}
\usepackage{booktabs}

\usepackage{amssymb}

\usepackage{lineno}

\biboptions{sort&compress}

\journal{Nuclear Instruments and Methods A}

\begin{document}

\begin{frontmatter}



\title{KRATTA, a versatile triple telescope array for charged reaction products}


\author{J.~\L{}ukasik\corref{cor1}}
\ead{jerzy.lukasik@ifj.edu.pl}
\cortext[cor1]{Corresponding author.}
\author{P.~Paw\l{}owski}
\author{A.~Budzanowski\fnref{fnd}}
\fntext[fnd]{Deceased.}
\author{B.~Czech}
\author{I.~Skwirczy\'nska}
\address{Institute of Nuclear Physics, IFJ-PAN, 31-342 Krak\'ow, Poland}
\author{J.~Brzychczyk}
\author{M.~Adamczyk}
\author{S.~Kupny}
\author{P.~Lasko}
\author{Z.~Sosin}
\author{A.~Wieloch}
\address{Institute of Physics, Jagiellonian University, 30-059 Krak\'ow, Poland}
\author{M.~Ki\v{s}}
\author{Y.~Leifels}
\author{W.~Trautmann}
\address{GSI, D-64291 Darmstadt, Germany}

\begin{abstract}

A new detection system KRATTA, Krak\'ow Triple Telescope Array, is presented.
This versatile, low threshold, broad energy range system has been built to
measure the energy, emission angle, and isotopic composition of light charged
reaction products. It consists of 38 independent modules which can be arranged
in an arbitrary configuration. A single module, covering actively about 4.5 msr
of the solid angle at the optimal distance of 40 cm from the target, consists of
three identical, 500 $\muup$m thick, large area photodiodes, used also for
direct detection, and of two CsI(1500 ppm Tl) crystals of 2.5 and 12.5 cm
length, respectively. All the signals are digitally processed. The lower
identification threshold, due to the thickness of the first photodiode, has been
reduced to about 2.5 MeV for protons ($\sim$65 $\muup$m of Si equivalent) by
applying a pulse shape analysis. The pulse shape analysis allowed also to
decompose the complex signals from the middle photodiode into their ionization
and scintillation components and to obtain a satisfactory isotopic resolution
with a single readout channel. The upper energy limit for protons is about 260
MeV.  The whole setup is easily portable. It performed very well during the
ASY-EOS experiment, conducted in May 2011 at GSI. The structure and performance
of the array are described using the results of Au+Au collisions at 400
MeV/nucleon obtained in this experiment.

\end{abstract}


\end{frontmatter}


\section{Introduction} 

Charged-particle detection and identification with isotopic resolution over a
large dynamic range in particle type and energy, is mandatory for studies of
isotopic effects in heavy-ion reactions. Phenomena related to the isotopic
composition of the reaction system and the emitted products have been shown to
be useful for exploring the properties of neutron-rich nuclear matter as, e.g.,
encountered in neutron stars [1,2]. Their importance will increase with the
availability of secondary beams of high intensity. Also studies of nuclear
structure near the limits of stability require increasingly sophisticated and
precise detection systems. A comprehensive overview of the recent and future
developments in this field of instrumentation can be found in Ref.
\cite{desouza}. 

Most of the existing charged particle detectors for the intermediate energy
range (up to a few tens or hundreds of MeV/nucleon) base their identification on
the two- or three-fold telescope method \cite{england, leo}. In order to provide
the lowest possible identification threshold, the first layer of the telescope
is usually made of a gas chamber (e.g. DELF \cite{DELF}, MULTICS \cite{MULTICS},
FASA \cite{FASA}, INDRA \cite{INDRA}, ISiS \cite{ISIS}, GARFIELD
\cite{GARFIELD}, FIASCO \cite{FIASCO}) or of a thin Si detector (e.g. FAUST
\cite{FAUST}, INDRA \cite{INDRA}, LASSA \cite{LASSA}, CHIMERA \cite{CHIMERA},
HiRA \cite{HIRA}, NIMROD \cite{NIMROD}, FAZIA \cite{FAZIA}). The first,
$\Delta$E, layer is then followed by one or two, thicker, Si detectors or
scintillators. 

The option with the silicon $\Delta$E layer, has the advantage of a better
resolution and is easier to handle, but usually results in higher thresholds and
is costly. The presented KRATTA modules belong to this class of telescopes,
however, they have been optimized to be budget friendly, without loosing the
quality of detection. Instead of using the Si detectors of different thickness,
they are using three identical, catalog size, photodiodes and two CsI(Tl) 
crystals. 
Thanks to the digital signal processing and the off-line pulse shape
analysis, the obtained mass resolutions for light charged particles are very
satisfactory in a broad energy range. On one hand, the pulse shape analysis
allowed to reduce the energy threshold, resulting from the relatively thick
first layer, by a factor of 3. On the other hand, it allowed to effectively
double the thickness of the silicon $\Delta$E layer, by combining the ionization
components from the first two photodiodes, and consequently, to improve the
resolution of the first segment of the telescope.

\section{Motivation and requirements} 

The main parameters of the KRATTA array have been motivated by the needs of the
ASY-EOS experiment \cite{asy-ref}. This experiment has been designed to study
the density dependence of the nuclear symmetry energy by measuring flows and
isotopic compositions of the reaction products from the $^{197}$Au+$^{197}$Au,
$^{96}$Ru+$^{96}$Ru, and $^{96}$Zr+$^{96}$Zr reactions at 400 MeV/nucleon.
During the experiment the most relevant products, neutrons and $Z$=1 and 2
particles, have been measured by the LAND \cite{land} detector and the direction
and magnitude of the impact vector were estimated using the CHIMERA
\cite{CHIMERA} and ALADIN ToF-Wall \cite{tof} detectors. The KRATTA array has
been designed to complement the neutron and hydrogen detection with LAND by
measuring the isotopic composition and flow of light charged reaction products
up to atomic number $Z \lesssim 7$, and specially, to identify the hydrogen and
helium isotopes with a resolution much better than achievable with LAND. The
array was placed on the opposite side of the beam with respect to LAND, and
covered approximately the same solid angle (160 msr). It has been designed to
detect energetic particles emerging from the ``mid-rapidity'' region of the 400
MeV/nucleon reactions. Modular design, portability, low thresholds (below 3
MeV/nucleon) and high maximum energy ($\sim$260 MeV/nucleon for $p$ and
$\alpha$), allow the array to be used in various configurations and experiments.
In particular, it will very well suit the needs of the future cyclotron facility
at IFJ-PAN in Krak\'ow, planned for proton beams from 70 to 250 MeV.  Last, but
not least, the KRATTA modules are also compatible with the design of the
existing Krakow Forward Wall Detector \cite{fwd} and can be used for reaching
complete coverage of the 2$\pi$ azimuthal angle.

\section{Active elements and geometry}

\begin{figure}[hbt] 
\centering
\includegraphics*[width=\linewidth]{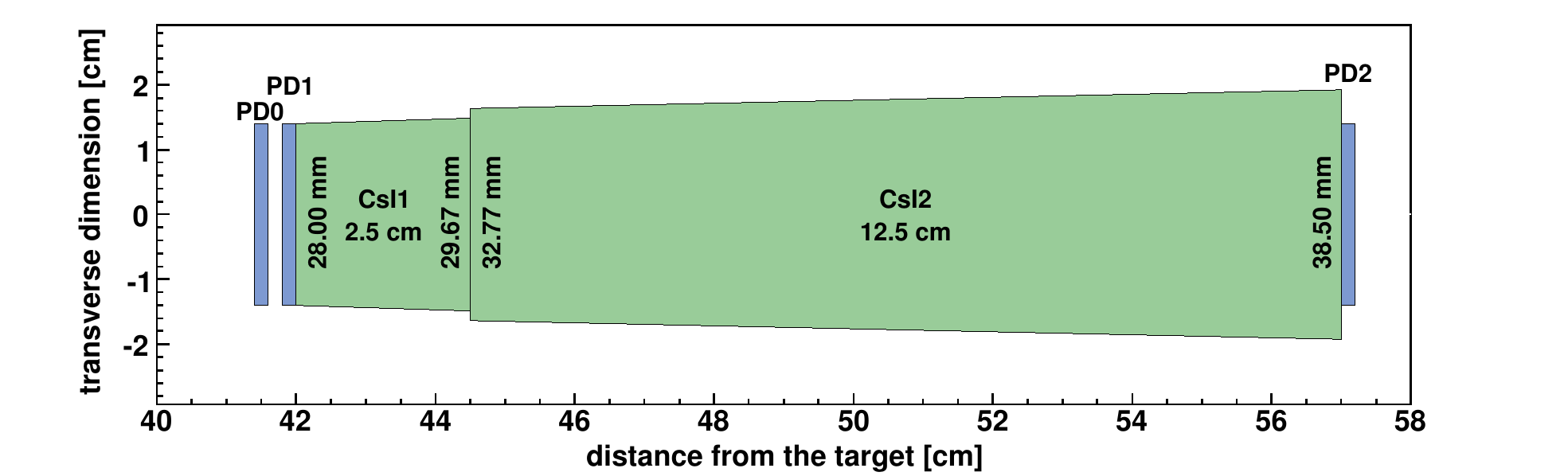}

\caption{Schematic layout of the active elements.}
\label{fig_mod} 
\end{figure}

The modules of KRATTA are composed of three large area HAMAMATSU PIN photodiodes
for direct detection \cite{pd-ref} and of two CsI(Tl) crystals \cite{csi-ref}.
The layout and dimensions of these active elements are presented in Fig.
\ref{fig_mod} and their main characteristics are summarized in Table \ref{t1}.

\begin{table} [hbt]
\begin{tabular*}{\linewidth}{@{\extracolsep{\fill}}ll}
\toprule
\multicolumn{2}{c}{Photodiodes \cite{pd-ref}}\\
Active Area$^{~a}$    &28$\times$28 mm$^2$\\
Thickness$^{~a}$      &500$\pm$15 $\muup$m  \\
Thickness non-uniformity$^{~a}$      &$<$ 3 $\muup$m  \\
Dead Layers$^{~a}$    &1.5 / 20 $\muup$m front / rear \\
Surface Orientation$^{~a}$    &(111) \\
Full Depletion Voltage$^{~b}$ (FD)  &120-135 V	 \\
Dark Current @ FD$^{~b}$   &6-16 nA, typ. 9 nA 	         \\
Terminal Capacitance @ FD$^{~b}$    &190$\pm$3 pF  	 \\
Rise Time (Laser $\deltaup$ pulse)$^{~a}$      &40 ns		 \\
\midrule
\multicolumn{2}{c}{CsI(Tl) Crystals \cite{csi-ref}} \\
Tl concentration$^{~c}$  &1500 ppm    \\
Light output non-uniformity$^{~c}$ &$<$ 7 \%    \\
Shape  	   &Truncated pyramids    \\
Tolerance	   &$\pm$0.1 mm \\
\midrule
\multicolumn{2}{c}{Wrapping \cite{esr-ref}} \\

  Reflectance$^{~a}$	    &$>$ 98 \%   \\
  Thickness$^{~a}$	    &65  $\muup$m  \\
\bottomrule
\end{tabular*}
\caption*{
$^{a}$ Values from technical note. \\
$^{b}$ Values from manufacturer's Inspection Sheet. \\
$^{c}$ Nominal values.}

\caption{Main characteristics of the active elements.}
\label{t1}
\end{table}

The first photodiode (PD0 in Fig. \ref{fig_mod}) serves as a Si $\Delta$E
detector providing the ionization signal alone. It has been ``reverse mount'',
i.e. the ohmic side towards the incoming particles.  The second photodiode
(PD1), naturally ``reverse mount'', works in a ``Single Chip Telescope'', SCT
\cite{sct-ref}, configuration and provides a composite signal combined of a
direct (ionization) component and of a scintillation component coming from the
thin crystal (CsI1). The third photodiode (PD2) reads out the light from the
thick crystal (CsI2) and, in addition, provides an ionization signal for
particles that punch through the crystal within its active area. 


\begin{figure}[!htb]
\centering
\includegraphics[width=\linewidth]{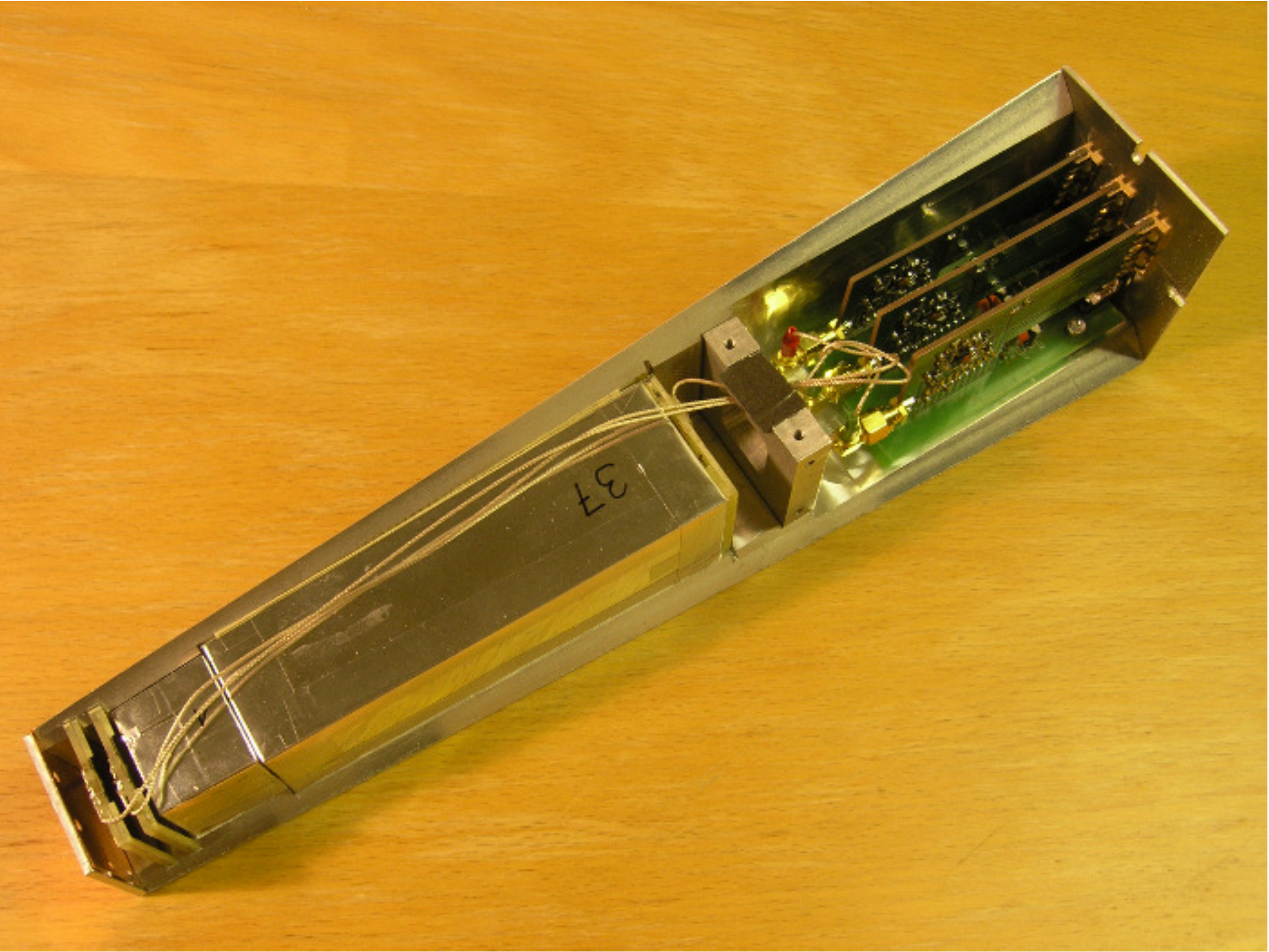}
\caption{Single module content.} 
\label{fig_mod_det1}
\end{figure}

\begin{figure}[!htb]
 \leavevmode
 \centering
  \includegraphics[width=\linewidth]{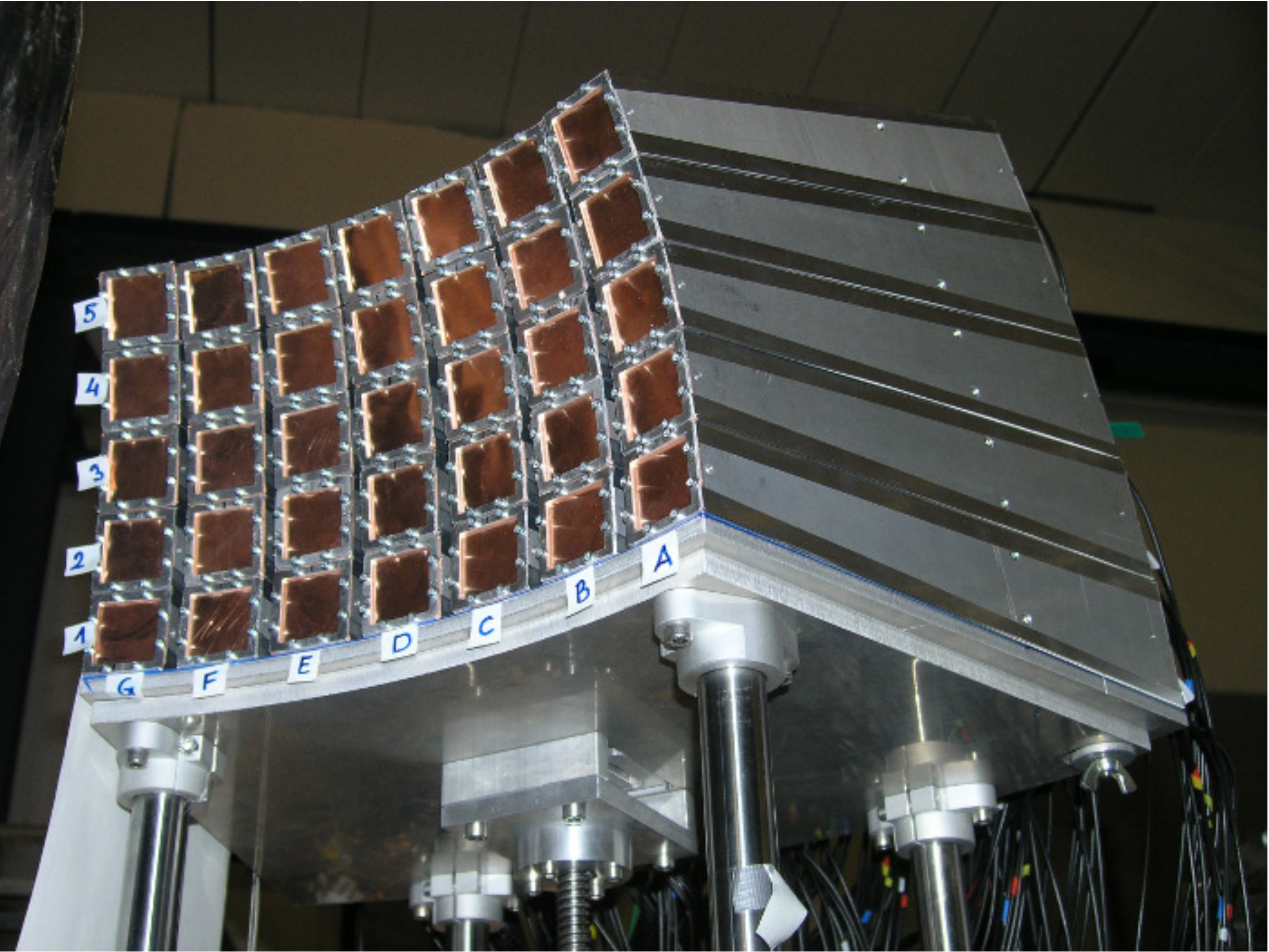}

  \caption{KRATTA in a 7$\times$5 configuration.} 

  \label{fig_mod_det2}

\end{figure}


\begin{table} [hbt]
\begin{center}
\begin{tabular*}{\linewidth}{r@{}l@{\extracolsep{\fill}}rrr}
\toprule
 \multicolumn{2}{c}{Fragment}& \multicolumn{1}{c}{E$_{low}$} & \multicolumn{1}{c}{E$_{int}$}  & \multicolumn{1}{c}{E$_{up}$}\\
\midrule
$^{1}  $&  H& 8.3  &  89.6&  254.4\\  
$^{4}  $& He& 8.3  &  89.4&  253.9\\  
$^{7}  $& Li& 9.5  & 103.6&  296.5\\
$^{20} $& Ne& 19.9 & 231.3&  719.0\\
$^{43} $& Ca& 26.7 & 339.7& 1134.2\\
$^{91} $& Zr& 34.0 & 513.9& 1911.8\\			     
$^{197}$& Au& 38.6 & 775.8& 3550.9\\
\bottomrule
\end{tabular*}

 \end{center}
 
\caption{Lower, E$_{low}$ and intermediate, E$_{int}$, thresholds and upper
limits, E$_{up}$, for selected species (in MeV/u). The thresholds correspond to
the energy losses in 500 $\muup$m of Si, in 1000 $\muup$m of Si + 2.5 cm of CsI
and in 1000 $\muup$m of Si + 15 cm of CsI, respectively. They have been
calculated using the ATIMA tables \cite{atima}.}

\label{t2}

\end{table}


The larger front face of the longer crystal with respect to the rear face of the
smaller one (see Fig. \ref{fig_mod}) permits its use at an about 2 times larger
distance from the target, in configuration that completes the missing half of
the FWD phoswich array \cite{fwd}. The crystals have been polished and wrapped
with a highly reflective ESR \cite{esr-ref} foil, except for the front and back
windows. The windows have been protected with 6 $\muup$m Mylar foils. The
crystals were optically decoupled. The photodiode chips have been glued onto
custom-made PCB frames and put in close optical contact with the crystal
windows. The active elements have been placed inside aluminum boxes together
with the charge preamplifiers (see Fig. \ref{fig_mod_det1}). The photodiode
frames and the aluminum housing reduced the geometric acceptance of a single
module to about 54\%. The active solid angle of a module amounts to 4.5 msr. The
entrance window has been made of a 100 $\muup$m thick copper foil. During the
experiment, 35 modules have been arranged in a 7$\times$5 array (Fig.
\ref{fig_mod_det2}), all placed at a radial distance of 40 cm from the target,
and operated in air. In a spherical coordinate system, with the origin at the
target and the beam axis defining the reference equatorial plane, the columns
follow the meridians from the beam level up to about 27$^{\circ}$ elevation
(latitude) and the rows span the azimuth angles (longitude) between 21$^{\circ}$
and 64$^{\circ}$ with respect to the beam direction. The energy thresholds
resulting from the thicknesses of the consecutive active layers are summarized
in Table \ref{t2}.  As will be shown later on, the low threshold can be further
reduced by applying a pulse shape analysis to the ionization signals of PD0. The
thresholds from Table \ref{t2} do not take into account the target, the air and
the entrance window present in front of the active layers. In the case of the
ASY-EOS experiment, these layers caused increase of the lower threshold,
E$_{low}$, by about 12.9 MeV/nucleon for protons and $\alpha$ particles (8.6,
1.8 and 2.5 MeV/nucleon energy losses due to the 350 $\muup$m Au target, 40 cm
of air and 100 $\muup$m of Cu, respectively). For low energy experiments, the
effects of air and the entrance window can be minimized by performing the
measurement in vacuum and removing the Cu foil or making it thinner.


\section{Electronics and data acquisition}

\begin{figure}[!htb]
 \centering
  \includegraphics[width=0.99\linewidth]{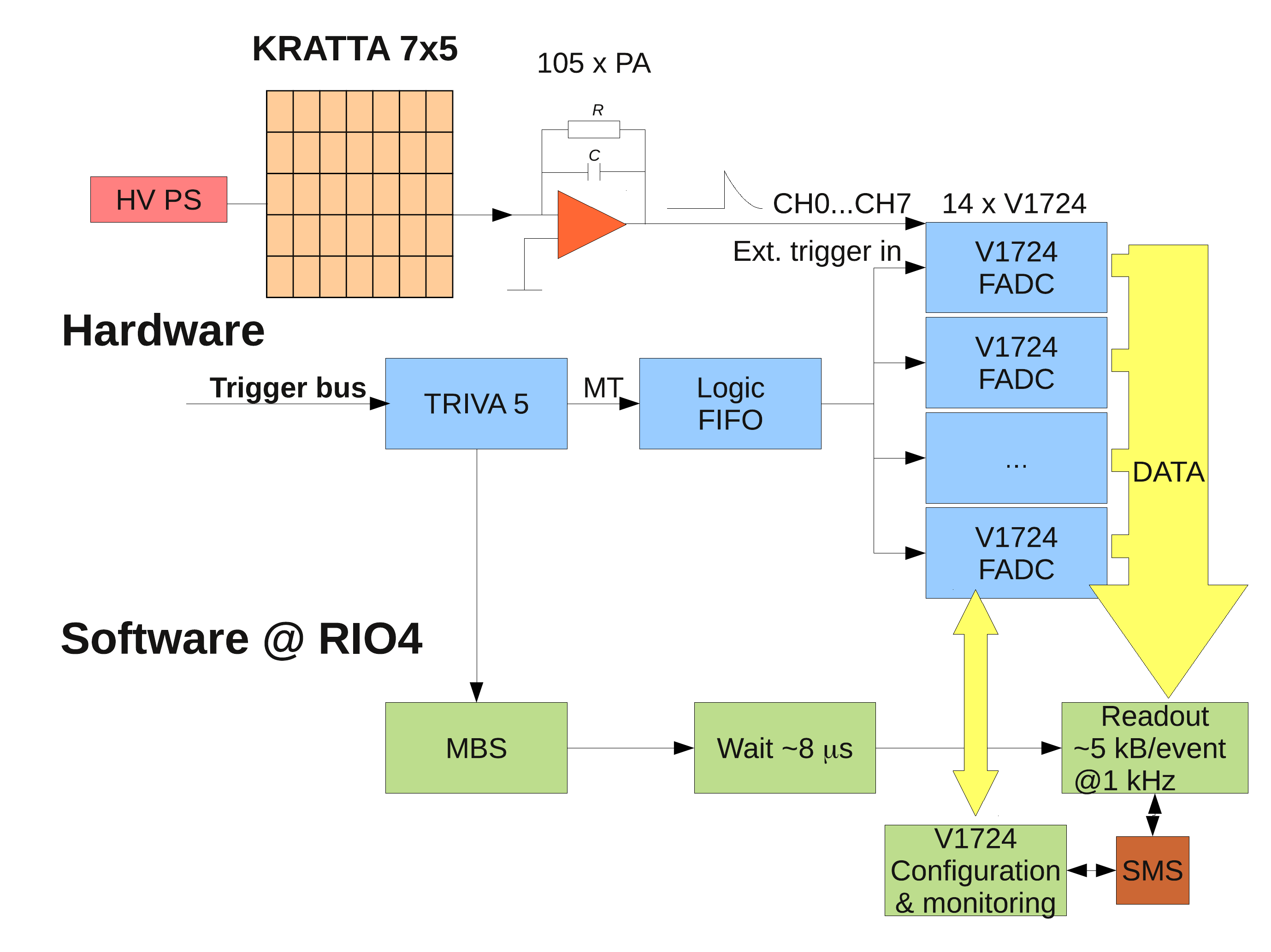}
  
  \caption{Analog, logical and digital signal flow chart. HV PS - high voltage
  power supply. PA - charge sensitive preamplifier. V1724 - CAEN digitizers.
  TRIVA5 - VME Trigger Synchronization Module \cite{triva-ref}. MT - master
  trigger. FIFO - logical Fan-In Fan-Out module. RIO4 - VME controller board
  \cite{rio-ref} running LynxOS. MBS - Multi Branch System, a GSI acquisition
  standard \cite{mbs-ref}. SMS - Shared Memory Segment.} 

  \label{fig_acq}

\end{figure}

The main electronics and data acquisition functions are presented schematically
in Fig. \ref{fig_acq}. The photodiodes (3 per module) have been reverse biased
at 120 V, using the in-house made 120-channel, remote controlled, high voltage
power supplies.  The signal from each photodiode has been integrated with the
own-design, low noise, charge preamplifier \cite{sosin}. 


The preamps were supplied with $\pm 6$ V and their dynamic range spanned about
3.6 V. Three nominal charge gains of the preamplifiers have been used, depending
on the azimuthal angle of the module: 44.5, 22.2 and 13.5 mV/MeV, with 1, 2 and
3.3 pF feedback capacitors, respectively. After optional amplification, the
signals have been digitized with the 100 MHz, 14 bit digitizers \cite{fadc-ref}
and stored for the off line analysis.  All logical and digital electronics
modules shown in Fig. \ref{fig_acq} have been controlled with the RIO4 board
\cite{rio-ref} within a single VME crate. During the experiment, the 14 Flash
ADC boards have been triggered with an external trigger split and delivered into
each FADC module. The stored waveforms spanned 5.12 or 10.24 $\muup$s (512 or
1024 time bins), with a 2 $\muup$s pre-trigger enabling a precise baseline
estimation. The shorter samples have been sufficient for the first photodiode
supplying the fast ionization signal alone. The expected data throughput
amounted to about 5 MB/s, assuming 1 kHz single hit rate. The actual data rate
did not go beyond this estimate during the experiment. The digitizers have been
remotely set up and monitored using a self-developed software. The data flow has
been controlled using the standard GSI MBS system \cite{mbs-ref}.

\section{Pulse shape analysis}

The pulse shape analysis has two main purposes in the case of KRATTA data. First
of all, it has to enable the decomposition of the signals from the middle
photodiode, PD1 (SCT), into the ionization and scintillation components.
Secondly, it is used to resolve masses of particles stopped in the first
photodiode, PD0, utilizing the relation between the range of a particle and the
time characteristics of the induced current signal \cite{pulse}.


The following assumptions have been made to accomplish these two goals. The
preamplifier response has been modeled using a simple parallel RC circuit
approximation \cite{knoll}: 
 \begin{minipage}[c]{0.24\linewidth}
 \leavevmode
 \centering
  \includegraphics[width=.99\linewidth]{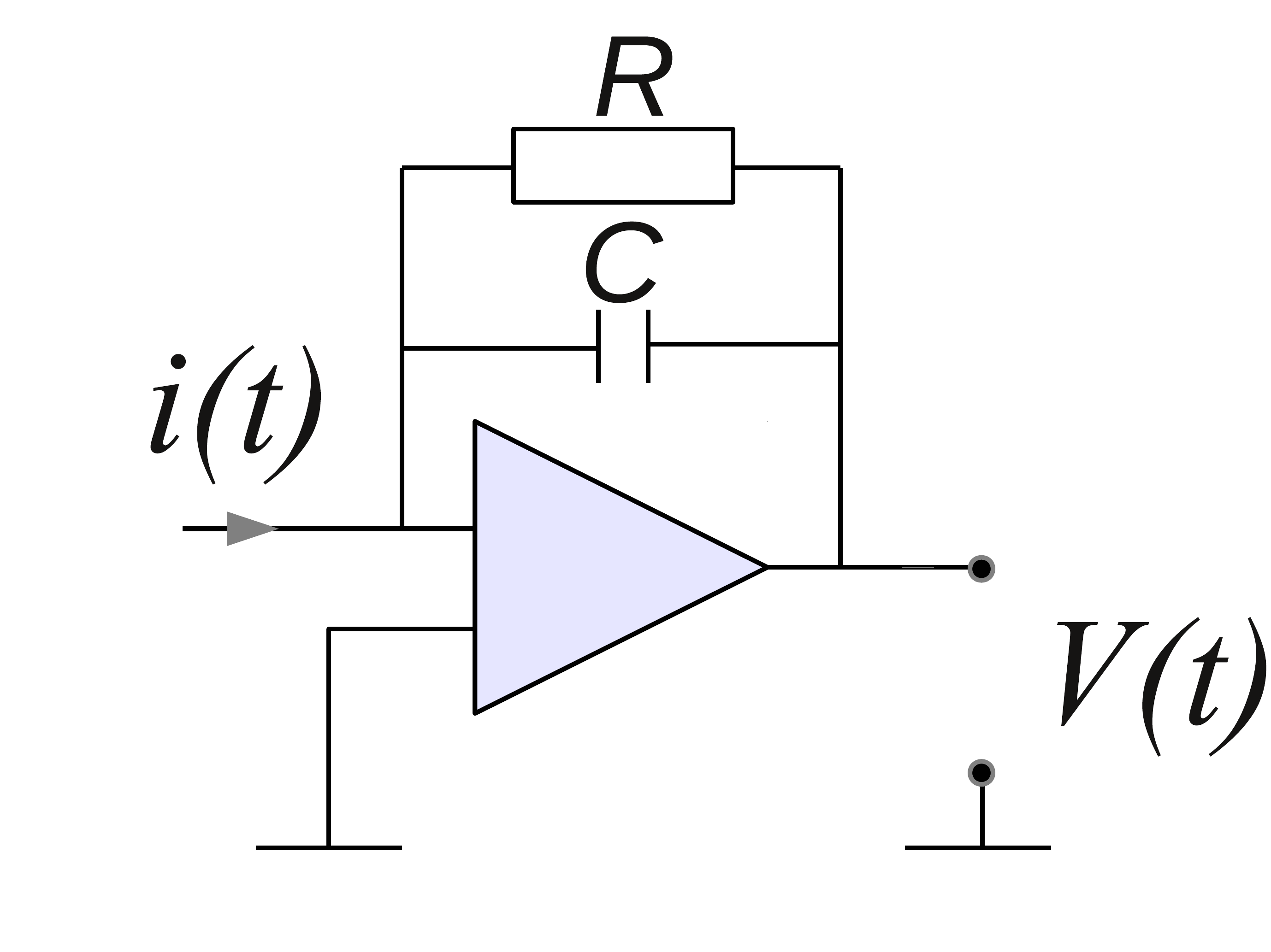}
 \end{minipage}
 \hfill
 \begin{minipage}[t]{0.70\linewidth}
  
\begin{equation}
 \frac{i(t)}{C}  = \frac{d V(t)}{dt} + \frac{V(t)}{RC}
  \label{eq_pa}
\end{equation}

  \label{fig_pa}

 \end{minipage}

\noindent where, $RC$ is the feedback coupling time constant, $i(t)$ is the
induced current due to the carrier motion in the photodiode, and $V(t)$ is the
measured voltage pulse. This relation assumes an infinite open loop gain, small
detector capacitance and a zero rise time of the charge integrator, which is an
idealization, but enables an analytical approach. The induced current has been
approximated with one direct (ionization), $i_{D}(t)$, and two scintillation,
$i_{Sk}(t)$, components, all of the same form: 
\begin{eqnarray}
\lefteqn{i_{D}(t) = Q_{D} \;\frac{e^{- t/\tau_{RD}} - e^{-t/\tau_{FD}}}{\tau_{RD} -
\tau_{FD}},}
\label{eq_d}\\
\lefteqn{i_{Sk}(t) = Q_{Sk} \;\frac{e^{- t/\tau_{RS}} - e^{-t/\tau_{Fk}}}{\tau_{RS} -
\tau_{Fk}},\quad k=1, 2}
  \label{eq_s}
\end{eqnarray}

\noindent where $Q$'s are the induced charges and the $\tau_{RD}, \tau_{RS}$ 
and the $\tau_{FD}, \tau_{Fk}$ are the rise and fall times for the respective
component. The assumed shapes attempt to account for both, the complicated
actual current pulse shape induced by the electrons and holes drifting in the
photodiode \cite{pulse}, and for the instrumental rise time of the preamp.  With
the assumed preamplifier response (\ref{eq_pa}) and the current shapes
(\ref{eq_d}) and (\ref{eq_s}), it was possible to obtain the corresponding
analytical model shapes of the waveforms, $V(t) = V_0 +  V_{D}(t) + V_{S}(t)$, with:
\begin{eqnarray}
\lefteqn{V_{D}(t) = RC \, Q_{D} \left ( 
\frac{e^{-\Delta t/RC}\, RC}{(RC - \tau_{RD}) (RC - \tau_{FD})}\right.} \nonumber \\ 
& & \!\!\!\!\!\!\!\!\!\!\!\!+ \left.\frac{e^{-\Delta t/\tau_{RD}}\, \tau_{RD}}{(\tau_{RD} - RC) (\tau_{RD} - \tau_{FD})}
 + \frac{e^{-\Delta t/\tau_{FD}}\, \tau_{FD}}{(\tau_{FD} - \tau_{RD}) (\tau_{FD} - RC)} \right )
  \label{eq_vd}\\
\lefteqn{V_{S}(t) = RC \sum_{k=1}^{2} Q_{Sk} \left ( 
\frac{e^{-\Delta t/RC}\, RC}{(RC - \tau_{RS}) (RC - \tau_{Fk})}\right.} \nonumber \\ 
& & \!\!\!\!\!\!\!\!\!\!\!\!+ \left.\frac{e^{-\Delta t/\tau_{RS}}\, \tau_{RS}}{(\tau_{RS} - RC) (\tau_{RS} - \tau_{Fk})}
 + \frac{e^{-\Delta t/\tau_{Fk}}\, \tau_{Fk}}{(\tau_{Fk} - \tau_{RS}) (\tau_{Fk} - RC)} \right )
  \label{eq_vs}
\end{eqnarray}

\noindent where, for generalization, $V_0$ is the baseline and $\Delta t =
t-t_0$, with $t_0$ being the beginning of the pulse.

The two scintillation components (\ref{eq_vs}) have been introduced to account
for the fast and slow decay modes of the CsI(Tl) crystals. The rise times,
$\tau_{RD}$ and $\tau_{RS}$, account for the photodiode, scintillator and the
preamp rise times. Overall, the model depends on 11 parameters listed in Table
\ref{t3}.

\begin{table} [hbt]
\begin{center}
\begin{tabular*}{\linewidth}{@{\extracolsep{\fill}}rlrl}
\toprule
 \multicolumn{2}{c}{Parameter}& \multicolumn{2}{c}{Value} \\
\midrule
 \multicolumn{4}{c}{Ionization} \\
$Q_{D}$   & Amplitude &      & fitted\\  
$\tau_{RD}$& Rise time	& $\sim$90 ns & fixed/fitted\\
$\tau_{FD}$& Fall time	& $<$300 ns & fixed/fitted\\
\midrule
 \multicolumn{4}{c}{Scintillation} \\
$Q_{S1}$   & Fast component amplitude	   &  & fitted\\
$Q_{S2}$   & Slow component amplitude	   &  & fitted\\  
$\tau_{F1}$& Fast fall time	   & $\sim$650 ns & fitted \\
$\tau_{F2}$& Slow fall time	   & $\sim$3.2 $\muup$s&  fixed\\
$\tau_{RS}$ & Rise time  	   & $\sim$140 ns &  fixed\\  
\midrule
 \multicolumn{4}{c}{Common} \\
$RC$       & Preamp fall time constant  & $\sim$6 $\muup$s &  fixed\\
$t_0$      & Time offset           & $\sim$2 $\muup$s & fitted\\			      
$V_0$      & Baseline              &  & fitted\\
\bottomrule
\end{tabular*}

 \end{center}
\caption{11 parameters of the model waveforms.}
\label{t3}
\end{table}

The preamp fall time constant parameter $RC$ has been determined
individually for each chip by selecting the pulses with the fast ionization
component alone. The resulting $RC$ constants were smaller than the nominal ones
by a few $\muup$s due to small leakage currents. In order to describe precisely
the shapes due to particles stopped in the first photodiode, PD0, both the time
constants, $\tau_{RD}$ and $\tau_{FD}$, were fitted and the scintillation
components were obviously not used. In case of PD1 and PD2, the rise and fall
times $\tau_{RD}$ and $\tau_{FD}$ were fixed. The fits were done using the
FUMILI \cite{fumili} minimization package, a relatively fast and precise
implementation of the Gauss-Newton algorithm. Constraining some of the
parameters was found inevitable, not only to obtain a good description of the
waveforms ($\,\chi^{2}$) but, at the same time, to maintain the global agreement
between the reconstructed amplitudes of the three components and the predictions
of the range-energy tables (see discussion of Figs. \ref{fig_raw01_atima},
\ref{fig_deco_atima}).

An additional advantage of using the digitization of the signals and the fitting
method, was the knowledge of the actual charges $Q$ for each component,
irrespectively of the substantial {\em ballistic deficit} (reduction of the
amplitude) due to a relatively short discharge time of the preamps ($RC\simeq$ 6
$\muup$s). For instance, the sum $Q_{S1}+Q_{S2}$ represents the total light
produced in the scintillators. For typical values of the parameters presented in
Table \ref{t3}, the ballistic deficit amounts to about 5-15\% for the ionization
signal and to about 25 and 50\% for the fast and slow CsI(Tl) components,
respectively. Usually, the fall time constant of the preamp is a compromise
between the level of pile-ups, the baseline variation, and the ballistic
deficit. However, since ballistic deficit is not an issue in our approach, the
short preamp fall times, of the order of the time span of the waveform and of
the slow CsI(Tl) decay time, have the additional advantage of making the maximum
and the tail of the pulse visible within the digitized sample.

Waveforms of low energy particles stopped in the first photodiode, PD0, have
been fitted  with a single component of the form (\ref{eq_vd}), with the time
constant parameters treated as free fit parameters. A more precise time
characteristics of the associated current pulse (\ref{eq_d}) was  needed to
perform a pulse shape based identification of these particles (see discussion of
Fig. \ref{fig_pd0}). 

\begin{figure}[!htb]
 \centering
  \includegraphics[width=\linewidth]{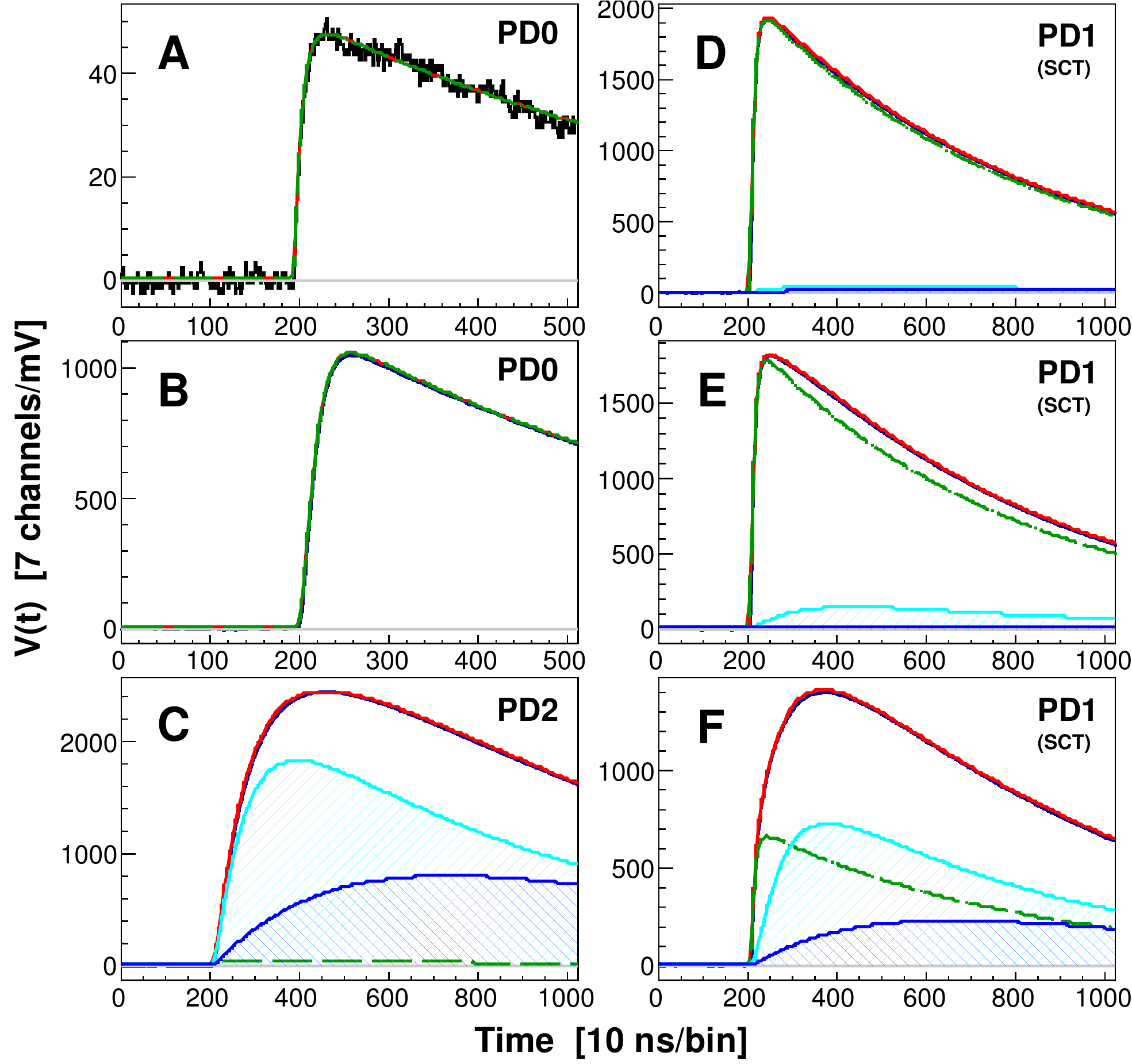}

  \caption{Example waveforms and their decomposition. Full time scale
  corresponds to 5 (panels A-B) or 10 (panels C-F) $\muup$s. 
  The ordinate represents 
  the measured voltage in the FADC channels.
  Histogram (when visible): measured waveform. 
  Solid line: sum of all fit components. 
  Dashed line: ionization component.
  Hatched areas: fast and slow CsI(Tl) scintillation components.
  A: low-energy electron, particle or gamma ray, barely above the acquisition 
     threshold, stopped in PD0 (see corresponding hit in Fig. \ref{fig_pd0}).
  B: $\alpha$ particle stopped in PD0 (see Fig. \ref{fig_pd0}).
  C: $\alpha$ particle stopped in CsI2 (see Fig. \ref{fig_raw12}, \ref{fig_rl2}).
  D: $\alpha$ particle stopped in PD1 (see Figs. \ref{fig_raw01}, 
   \ref{fig_sct01s}, \ref{fig_sct01d},\ref{fig_good_res}). 
  E: $\alpha$ particle barely punching through PD1 and stopped in CsI1 
  (see Figs. \ref{fig_raw01}, \ref{fig_sct01s}, \ref{fig_rl}, 
  \ref{fig_good_res}). 
  F: higher energy $\alpha$ particle stopped in CsI1 
  (see Figs. \ref{fig_raw01}, \ref{fig_sct01s}, \ref{fig_rl}, 
  \ref{fig_good_res}.)}

  \label{fig_puls1}

\end{figure}

The quality of the fits is demonstrated in Fig. \ref{fig_puls1}. The ordinate
represents the measured voltage, in the FADC channels, which is about twice
smaller than the actual one due to the matching of the 50 $\Omega$  FADC input
impedance. The hits corresponding to the selected waveforms are also marked in
the identification maps presented in the next section.

Panel A of Fig. \ref{fig_puls1} shows a waveform for a low energy electron,
particle or gamma, registered barely above the acquisition threshold and stopped
in PD0 (see also Fig. \ref{fig_pd0}). Here the measured histogram is visible and
the deviations from the smooth fit visualize the level of the total noise. The
resultant signal distortions, including the photodiode, preamplifier, FADC and
pickup noise sources, have been observed on the level of 0.4 mV rms,
corresponding to about 30 keV.

Panel B of Fig. \ref{fig_puls1} presents the quality of the fit with the shape
given by an ionization component alone (\ref{eq_vd}), applied to particles
stopped in PD0. Its location in the identification map is presented in Fig.
\ref{fig_pd0}. This fit allowed for derivation of both, the rise and fall times
and, therefore, also of the {\em mode} (position of maximum) of the associated current
signal (\ref{eq_d}):
\begin{equation}
mode = \frac{\tau_{RD}\; \tau_{FD}}{\tau_{RD} -\tau_{FD}} \log{\frac{\tau_{RD}}{\tau_{FD}}}
  \label{eq_m}
\end{equation}

Panel C of Fig. \ref{fig_puls1} shows the waveform of a high energy $\alpha$
particle stopped in the CSI2 crystal. The corresponding hit has been marked in
Figs. \ref{fig_raw12} and \ref{fig_rl2}.

Panels D-E-F of Fig. \ref{fig_puls1} show the evolution of the pulse shape of an
$\alpha$ particle detected and stopped in the SCT as its energy increases. The
corresponding hits have been marked, when possible, in Figs. \ref{fig_raw01},
\ref{fig_sct01s}-\ref{fig_good_res}.

The fitting of signals from PD2 (Fig. \ref{fig_puls1}, panel C) produces small
artificial ionization contributions for particles which actually do not hit the
photodiode (see Fig. \ref{fig_raw12}). It amounts, on average, to 1.7$\pm$0.5\%
of the total amplitude. Correspondingly, the artificial contribution of a
scintillation component for particles stopped in the PD1 photodiode, and thus
producing no light (see representative hit in panel D), is about 3.8$\pm$0.6\%. 
These numbers specify the quality of the pulse shape parametrization and the
systematic uncertainty of the decomposition into different components. The
artificial scintillation component has been removed by subtracting its well
defined fraction from the total amplitude and thus making the ionization and
scintillation components of the SCT almost perfectly orthogonal.

\section{Performance}

\begin{figure}[!htb]
 \centering
  \includegraphics[width=\linewidth]{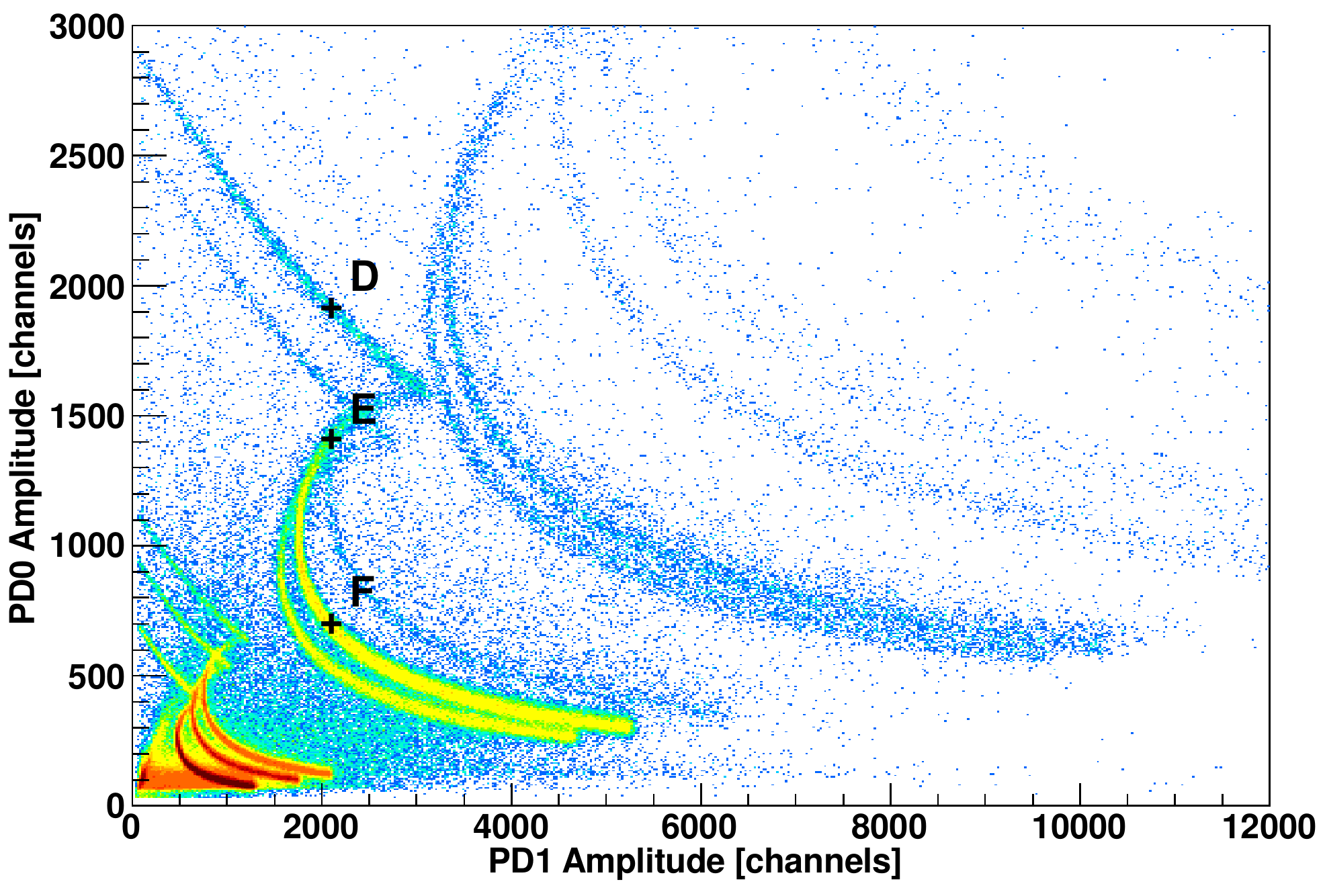}
  \caption{$\Delta$E-E ID map for the first two photodiodes, 
  PD0 vs PD1(SCT), for particles stopped in PD1 or in the thin crystal (CsI1).} 
  \label{fig_raw01}
\end{figure}

\begin{figure}[!htb]
 \centering
  \includegraphics[width=\linewidth]{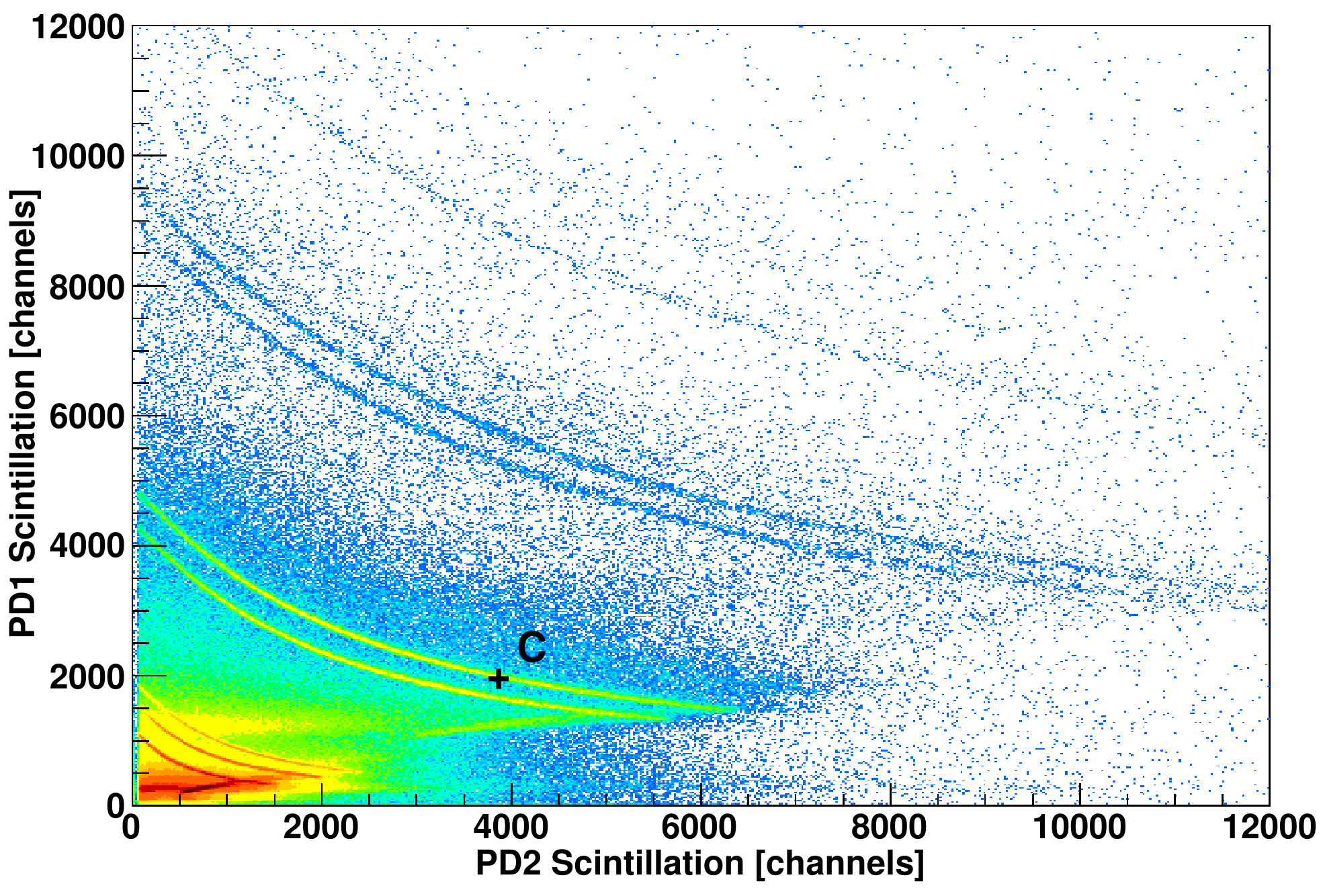}

  \caption{$\Delta$E-E ID map of scintillation signals for CsI1 vs CsI2
  detected by PD1 and PD2.} 

  \label{fig_raw12}

\end{figure}

\begin{figure}[!htb] 
\centering
 \includegraphics[width=\linewidth]{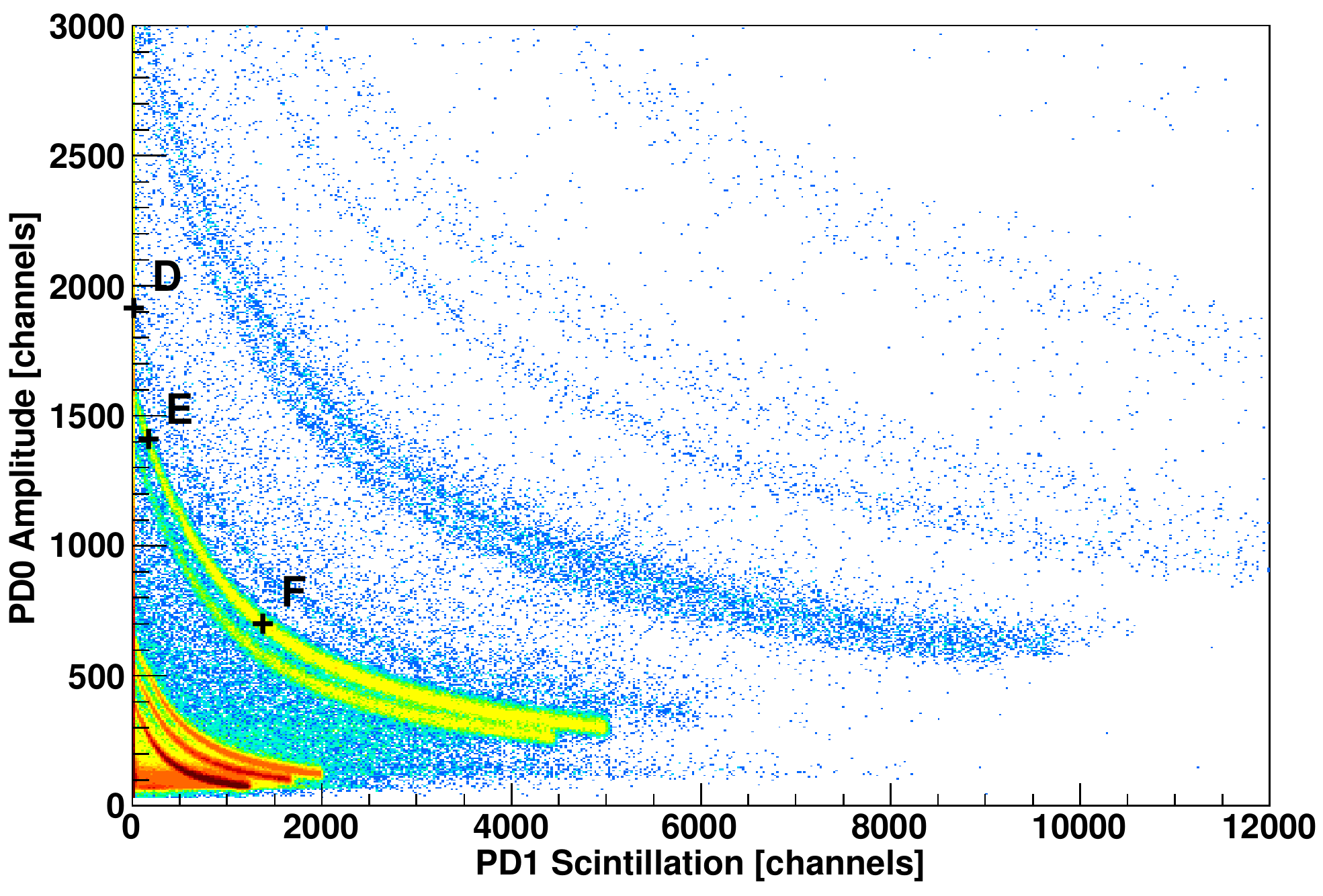}

 \caption{Decomposition of the map from Fig. \ref{fig_raw01}. PD0 vs
 scintillation detected by PD1 - for particles punching through PD1 and stopped
 in CsI1.}  

  \label{fig_sct01s}

\end{figure}

\begin{figure}[!htb] \centering
 \includegraphics[width=\linewidth]{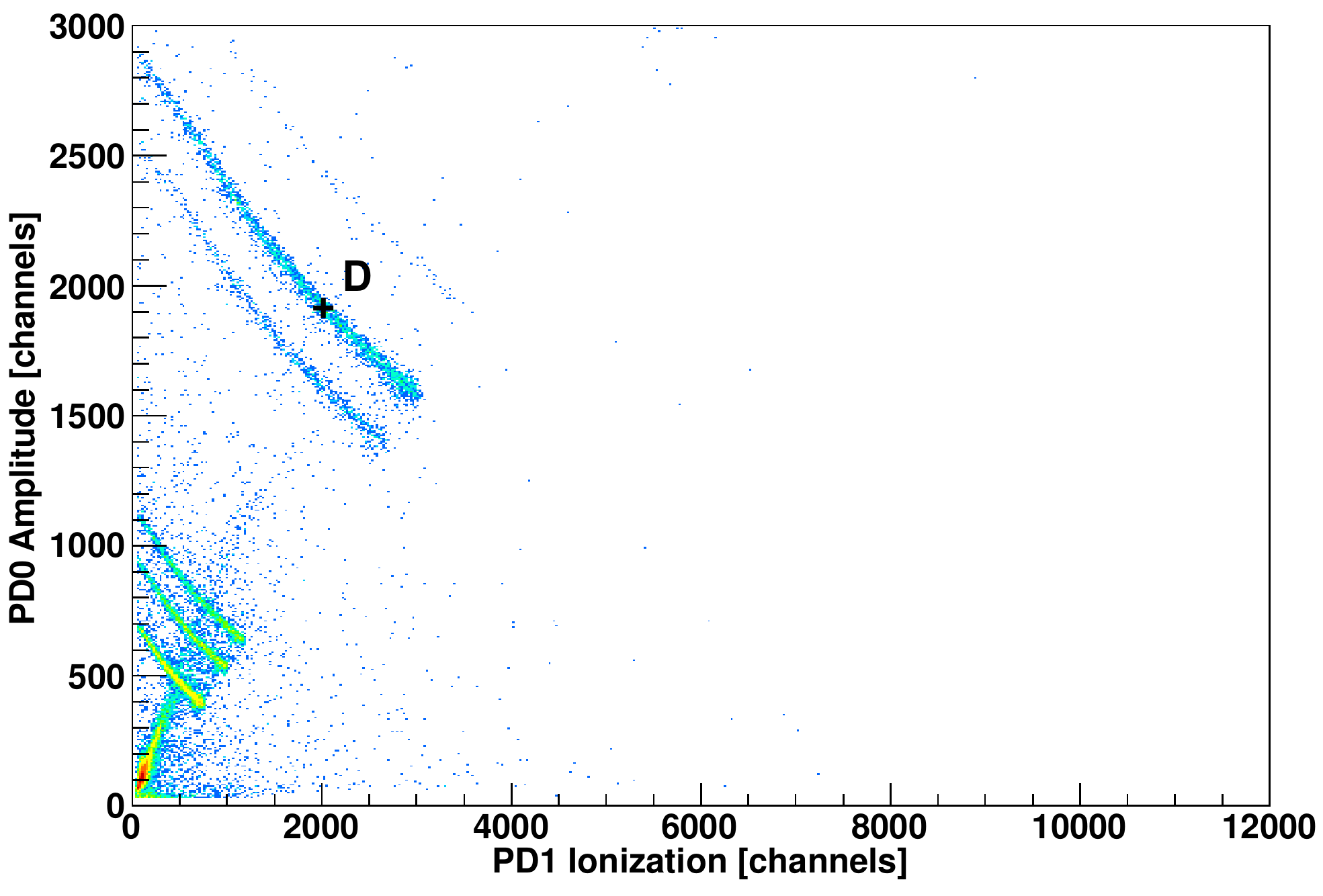} 
 
 \caption{Decomposition of the map from Fig. \ref{fig_raw01}. PD0 vs ionization
 signal from PD1 - for particles stopped in PD1 (producing no light).}  

  \label{fig_sct01d}

\end{figure}

\begin{figure}[!htb]
 \centering
  \includegraphics[width=\linewidth]{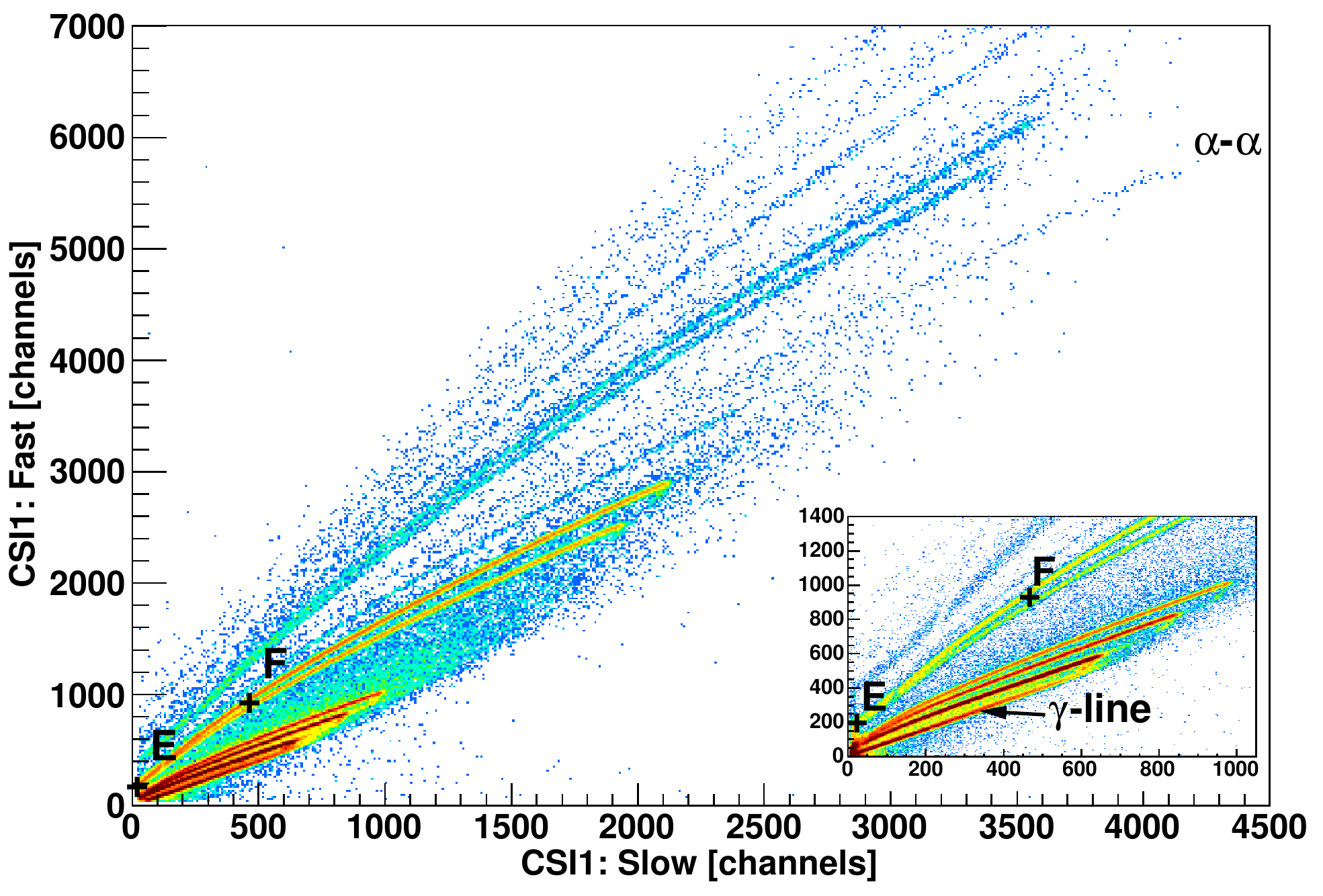}

  \caption{Fast vs Slow components of the scintillation in CsI1. The inset shows
more clearly a $\gamma-$line and the $p,\, d,\, t$ lines. } 

  \label{fig_rl}

\end{figure}

\begin{figure}[!htb]
 \centering
  \includegraphics[width=\linewidth]{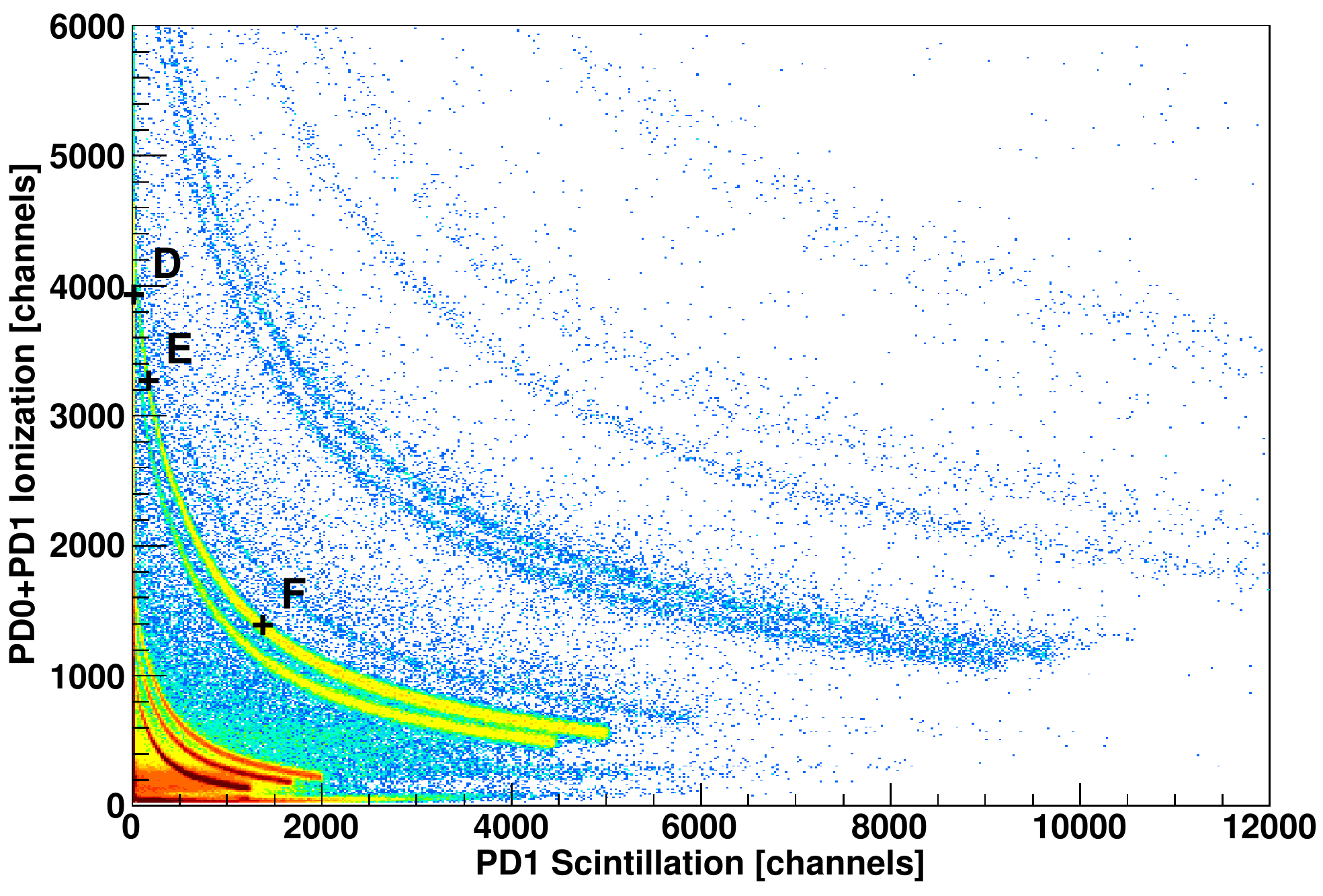}
  
  \caption{Summed PD0+PD1 ionization components vs scintillation in PD1 for
  particles stopped in CsI1. Note a slightly improved resolution compared to
  Fig. \ref{fig_sct01s} (see inset in Fig. \ref{fig_z})} 

  \label{fig_good_res}

\end{figure}

\begin{figure}[!htb]
 \centering
  \includegraphics[width=\linewidth]{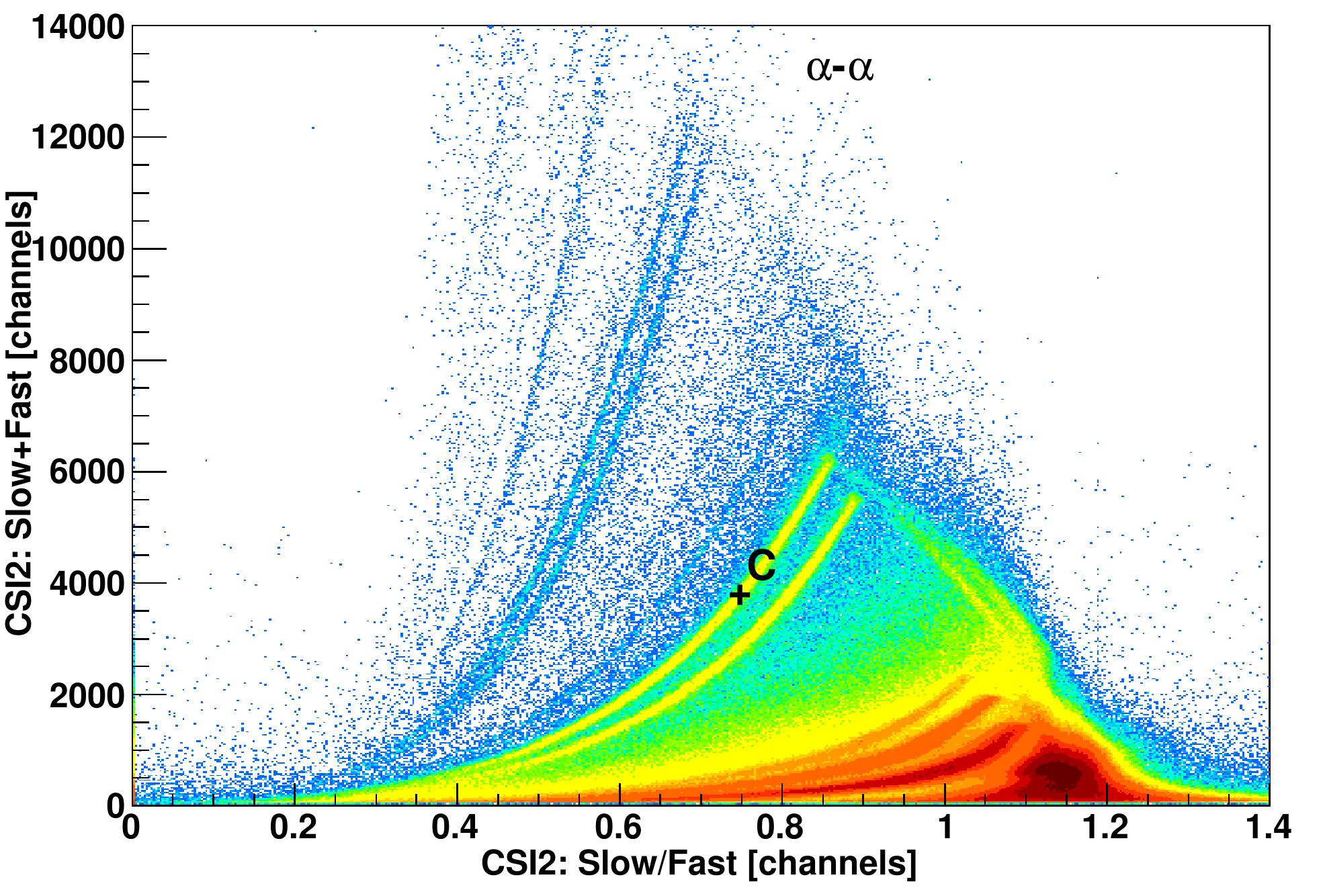}

  \caption{Total light vs Slow over Fast components of the scintillation in
CsI2. Note, that unlike in Fig. \ref{fig_raw12}, the punch through lines do
not cross the lower lying isotope lines.} 

  \label{fig_rl2}

\end{figure}

\begin{figure}[!htb]
 \centering
  \includegraphics[width=\linewidth]{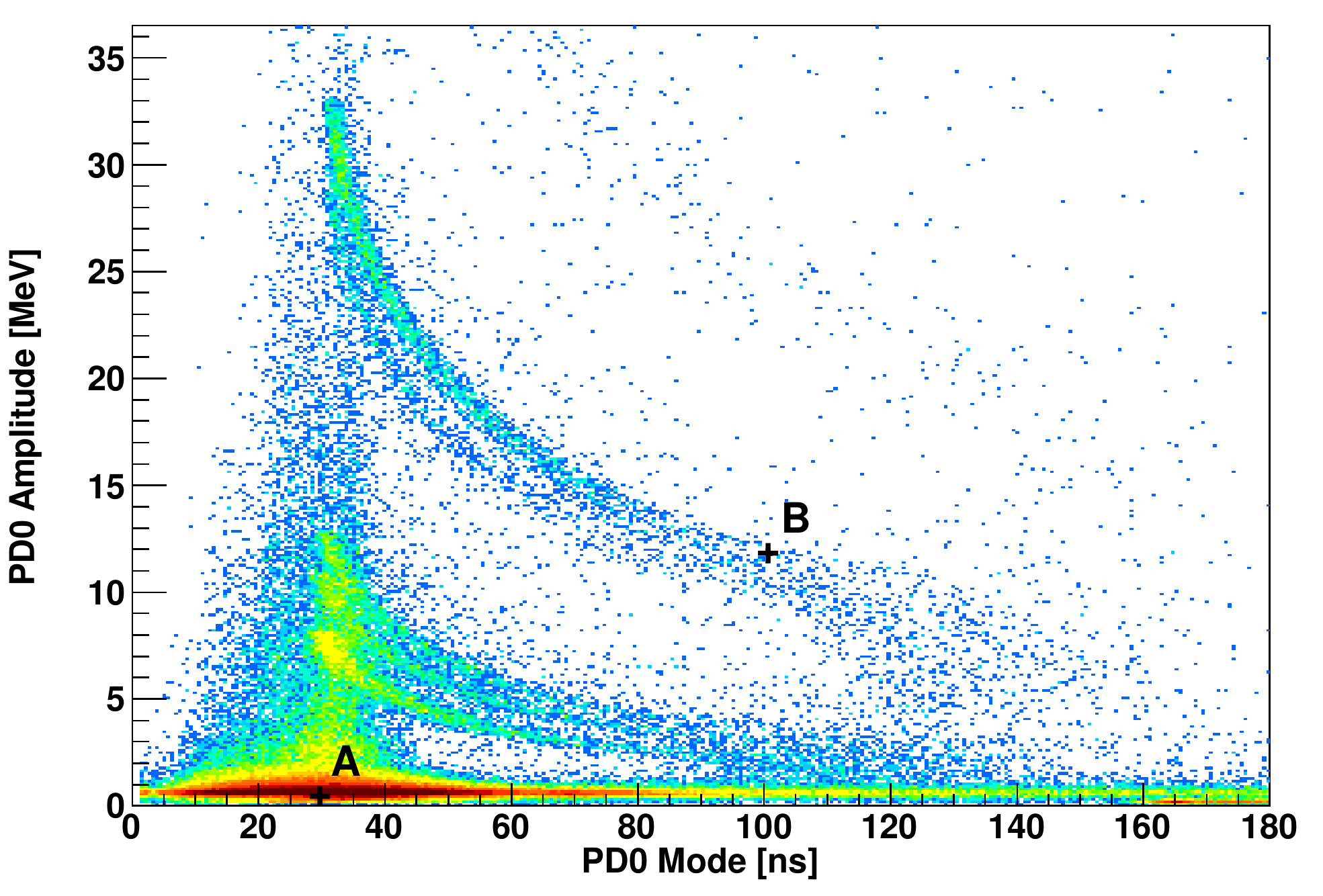}

  \caption{Amplitude vs mode of the current signal for particles stopped in
  PD0, obtained from a single Si chip.} 

  \label{fig_pd0}
\end{figure}

Figures \ref{fig_raw01}-\ref{fig_pd0} present various identification (ID) maps
obtained by using the parameters of the reconstructed waveform components.
The strongest lines in Figs. \ref{fig_raw01}-\ref{fig_pd0} correspond to  $p$,
$d$, $t$, $^{3}He$, $\alpha$, and so on, from bottom to top, respectively (see
Fig. \ref{fig_raw01_atima} for precise labeling). 

The reconstructed amplitude maps for the first two and the last two photodiodes
of the KRATTA module are presented in Figs. \ref{fig_raw01} and
\ref{fig_raw12}. 

The identification map in Fig. \ref{fig_raw01} shows a complex spectrum with
each ``ID-line'' composed of two parts: an ordinary Si-Si hyperbolic segment at
low energies, for particles stopped in PD1, and a more curved part for particles
punching through the PD1 photodiode and stopped in the CsI1 crystal. Due to line
crossing and its complex structure, this kind of a map cannot be used for
identification, however thanks to the presence of many characteristic
punch-through points and curvatures, it is very well suited for energy
calibration purposes.

Fig. \ref{fig_raw12} shows a $\Delta$E-E  map for scintillation signals detected
by a second vs third photodiode. Apart from the very good isotopic resolution,
one can see a substantial background from the secondary reactions/scatterings in
the crystals and also from particles crossing the module at an angle and not
originating from the target, as well as back-bendings corresponding to particles
punching through the thick crystal (see discussion at the end of this section).
Thanks to the pulse shape decomposition, the ionization component for particles
hitting the photodiode {\em(nuclear counter effect)} at the back of the thick
crystal has been easily removed.

Using the individual reconstructed amplitudes for the ionization and for the
scintillation components, the map of Fig. \ref{fig_raw01} can be decomposed into
the PD0-PD1(Si) and PD0-PD1(CsI) components of the SCT segment (Figs.
\ref{fig_sct01d} and \ref{fig_sct01s}). This makes effectively KRATTA even a
four-fold telescope.

\begin{figure}[!htb] \centering
\includegraphics[width=\linewidth]{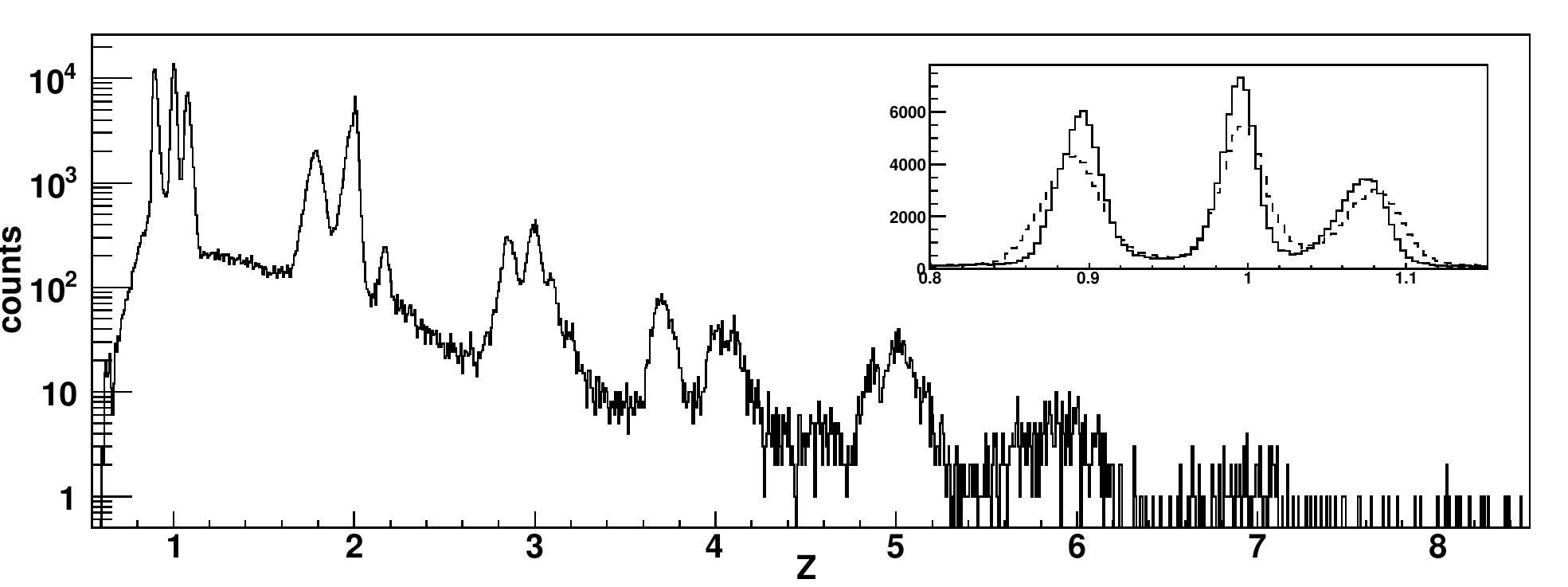} 

\caption{Particle identification spectrum (charge distribution) obtained from
the map of Fig. \ref{fig_good_res}, for particles stopped in the thin crystal.
The inset shows the $p,\, d,\, t$ peaks (in linear scale) obtained from the Fig.
\ref{fig_good_res} (solid) and Fig. \ref{fig_sct01s} (dashed histogram),
respectively.}   

\label{fig_z}
\end{figure}

The isotopic resolution visible from Fig. \ref{fig_sct01s} can be improved by
summing up the reconstructed ionization components from PD0 and PD1, and thus,
increasing the effective thickness of the first $\Delta$E layer to 1 mm of Si.
See also inset in Fig. \ref{fig_z}. Fig. \ref{fig_good_res} presents a classical
$\Delta$E-E ID map obtained from Fig. \ref{fig_raw01} thanks to the pulse shape
decomposition (see discussion of Fig. \ref{fig_deco_atima} for more details on
this transformation and Fig. \ref{fig_z} for the corresponding mass resolution.)

Figures \ref{fig_rl} and \ref{fig_rl2} show the ``Fast-Slow'' ID maps for the
CsI1 and CsI2 crystals, respectively. The latter represents a variant of the
standard representation, using the total light vs the ratio Slow over Fast (see
e.g. \cite{benrachi}). The Fast-Slow representation is very useful in many
respects in addition to the standard $\Delta$E-E one: One can observe a clear
double alpha line ($\alpha$-$\alpha$ in Figs. \ref{fig_rl} and \ref{fig_rl2})
which is hidden in the $\Delta$E-E representation behind the Li isotope lines.
The Fast-Slow map shows a clear ``$\gamma$-line'' - a strong line below the
proton line due to $\gamma$-rays (see inset in Fig. \ref{fig_rl}), which can be
precisely isolated and removed from the $\Delta$E-E map. Last, but not least, as
can be seen from Fig. \ref{fig_rl2}, the punch through segments do not cross the
lower lying isotope lines in a way they do in  case of the $\Delta$E-E maps
(Fig. \ref{fig_raw12}). Thus, the Fast-Slow maps allow to isolate the punch
through segments in a much more precise way. Unfortunately, the punch through
segments eventually merge with the $\gamma$-lines, which makes it difficult to
discriminate between these two.

Figure \ref{fig_pd0} demonstrates yet another powerful feature of the pulse
shape analysis. It allowed to perform an identification of the majority of light
particles stopped in the first photodiode by plotting the amplitude vs mode of
the reconstructed current signal (\ref{eq_m}). Due to the constant value of the
field within the intrinsic region of the PIN diode the enhancement of the
resolution for stopped particles is not expected to be as pronounced as in the
case of reverse mount PN detectors \cite{reverse}, nevertheless, the relation
between the range and the carrier collection time seems to be still strong
enough to enable the isotopic separation of light particles. This method allowed
to reduce the lower identification threshold, due to the thickness of the first
photodiode,  (see Table \ref{t2}) from 8.3 to about 2.5 MeV for protons, where
they are still resolvable from deuterons (see two bottom lines in Fig.
\ref{fig_pd0}). This effectively corresponds to the reduction of the thickness
of the first $\Delta$E layer from 500 to about 65 $\muup$m of Si. 

One should stress, that the time scale presented in Fig. \ref{fig_pd0} exceeds
by about a factor of 3 the collection times estimated on the basis of the
mobilities of the carriers alone, thus the plasma delay, or other effects, seem
to be quite important in slowing down the collection process. 

\begin{figure}[!htb] \centering
\includegraphics[width=\linewidth]{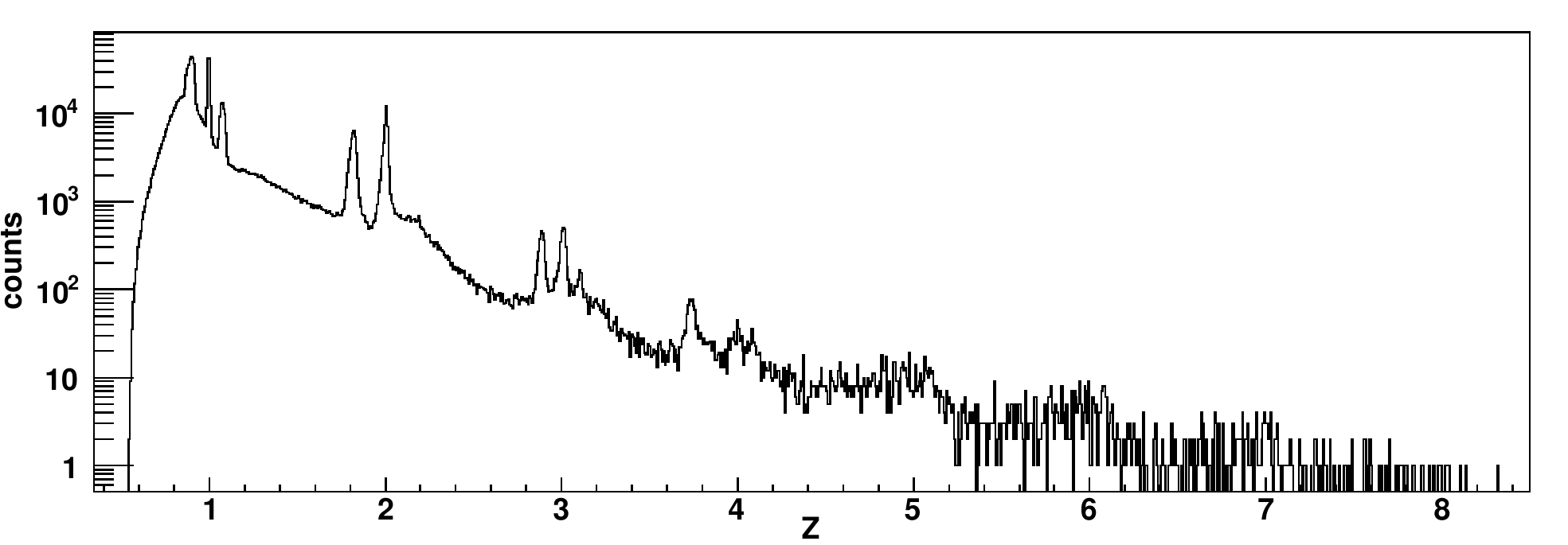} 

\caption{Same as Fig. \ref{fig_z}, but obtained from the map of Fig.
\ref{fig_raw12}, for particles stopped in the thick crystal.}  

\label{fig_z_thick}
\end{figure}

The mass resolution for particles stopped in the thin crystal (CsI1) and in the
thick one (CsI2) can be viewed from Figs. \ref{fig_z} and \ref{fig_z_thick},
respectively.

The background in Fig. \ref{fig_z}, resulting mainly from the secondary
reactions/scattering in the crystal, amounts to about 6\% for the hits below the
$p,\, d,\, t$ peaks. For energetic $Z=1$ particles traversing the thick crystal
(12.5 cm of CsI) the secondary reaction probability amounts to about 46\% (Fig.
\ref{fig_z_thick}). These probabilities agree reasonably well with the simple
estimates obtained using the nuclear collision length of 22.30 cm for CsI
\cite{pdg}, which yields the probabilities of 11\% and 43\% for 2.5 and 12.5 cm
of CsI, respectively. The background measured under the $p,\, d,\, t$ lines
includes also some contribution from the secondary reactions of neutrons and
heavier charged particles, as well as from the accidental coincidences. The
$\gamma$ and punch-through hits have been removed from the background, in both
cases. The high secondary interaction probability obviously deteriorates the
identification quality and defines the limits for applying the telescope method
to intermediate energies.

\section{Discussion and remarks}

\begin{figure}[!htb]
 \centering
  \includegraphics[width=\linewidth]{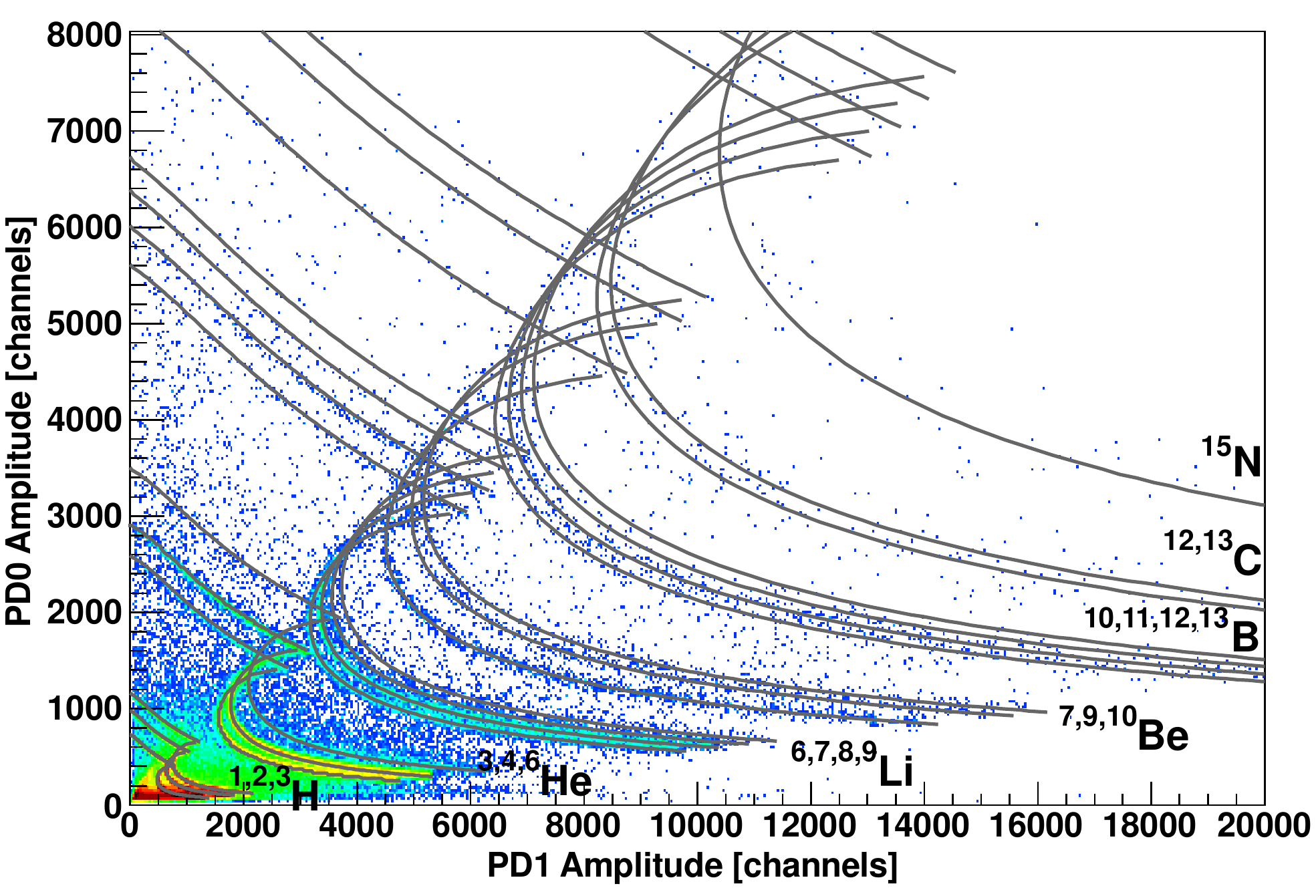}
  \caption{Same as Fig. \ref{fig_raw01} but with the superimposed ID lines 
  calculated using the ATIMA range-energy tables.} 
  \label{fig_raw01_atima}
\end{figure}

Since the parametrization of the pulse shapes is only approximate and has no
deep physical background, one can ask how precise is the decomposition into
individual components and how physical they are. In order to address these
questions, the $\Delta$E-E map of the SCT (PD1+CsI1) has been compared to the
predictions of the ATIMA range-energy tables. Such a comparison requires the
knowledge of the energy calibration of both, the silicon photodiodes and of the
CsI(Tl) light output. The calibration has been performed using the ID map of
Fig. \ref{fig_raw01} (see Fig. \ref{fig_raw01_atima}) which is relatively
insensitive to the details of the decomposition and is perfectly suited for the
energy calibration due to its richness in characteristic punch-through
points and curvatures.

The calibration routine allowed to adjust the thicknesses of the dead and active
layers, the energy calibration parameters, as well as the light-energy
conversion parameters. For the latter, a simple integrated Birks' formula (see
e.g. \cite{knoll}) has been used, with $dE/dx \propto A\, Z^{2}/E$, yielding a
commonly applied two-parameter Light-Energy relation \cite{horn}, applicable for
particles stopped in the scintillator: 
\begin{equation} 
\mbox{Light} = a_{1} \left (E - a_{2}\, A\, Z^{2} 
\log{\left [ 1+\frac{E}{a_{2}\, A\, Z^{2}}\right ]} \right )
\label{eq_birks} 
\end{equation}
with $E$ being the energy,  $A, Z$ - mass and atomic numbers of the stopped
particle, $a_{1}, a_{2}$ - the gain and quenching parameters.

\begin{figure}[!htb] \centering
\includegraphics[width=\linewidth]{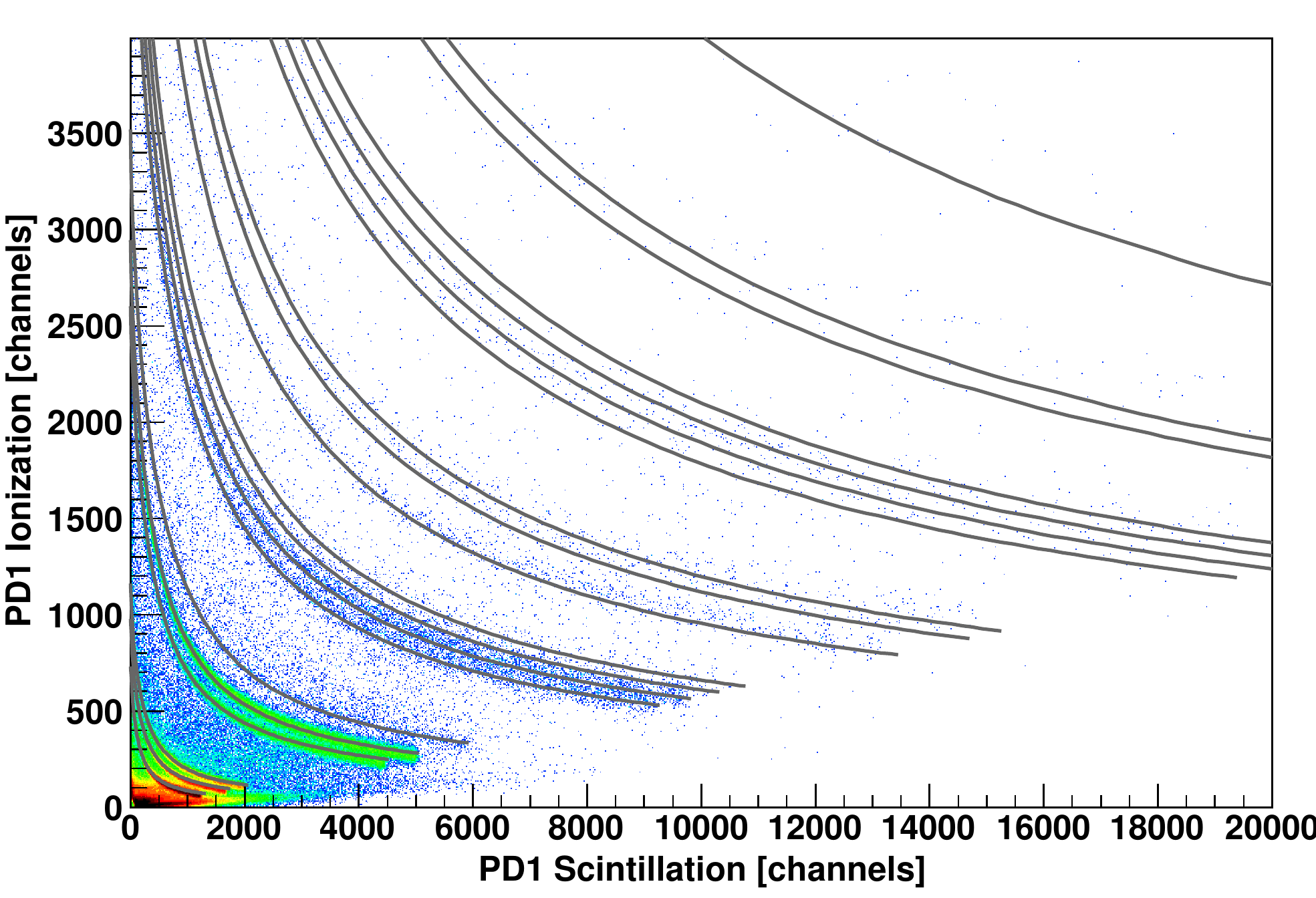} 

\caption{Decomposed SCT $\Delta$E-E identification map (obtained with a single
readout channel) with the  superimposed ID lines calculated using the ATIMA
tables. The sequence of lines is the same as in Fig. \ref{fig_raw01_atima}. }  

\label{fig_deco_atima}
\end{figure}

Having this (inverse) calibration it was possible to superpose the ATIMA lines
on the decomposed ID map for the Single Chip Telescope segment of the KRATTA
module. The result is presented in Fig. \ref{fig_deco_atima} and is worth a
comment.

The observed very good agreement between the reconstructed amplitudes and the
range-energy table predictions is quite non-trivial taking into account the
simplicity of the pulse shape parametrization. In order to obtain this kind of
agreement, which is still a compromise, it was not enough to set all the
parameters free and search for the minimum of the $\,\chi^{2}$ distribution.
Such an automatic procedure guarantees the best fit to the waveforms, but not
necessarily guarantees the follow up of the trends defined by the range-energy
tables. In general, an automatic procedure can lead, depending also on the
starting values, to some local minima, produce discontinuities or lead in
directions diverging from the ATIMA lines. The presented agreement has been
obtained after searching for the optimal values and fixing some of the
parameters (see Table \ref{t3}). The agreement with the ATIMA lines could be
made even better, for instance by freeing the $\tau_{RS}$ parameter, however at
the expense of loosing completely the mass resolution in Fast-Slow
representation (contrary to the case presented in Fig. \ref{fig_rl}). Thus, the
presented agreement is a result of an iterative procedure of fixing some of the
parameters and also of the energy calibration parameters, but once the crucial
parameters are constrained the fitting proceeds automatically.

The whole procedure could probably be better automatized by using more realistic
pulse shape parametrizations, especially for individual electron and hole
components of the ionization signal, taking into account the plasma delay
effects, interactions between carriers, diffusion effects, etc \cite{pulse}.
However, to our knowledge, such pulse shape parametrizations are still to be
developed. Definitely, a more realistic preamplifier response function would
improve the resolution and would allow for disentangling between physical and
instrumental parameters. A more realistic parametrization and analysis would be
worth the effort of implementing it to e.g. better understand the relation
between the range and the charge collection times which enable the
identification of the particles stopped in the silicon detectors. Nevertheless,
any more sophisticated analysis would definitely slow down even more the
analysis.

\begin{figure}[!htb] \centering
\includegraphics[width=\linewidth]{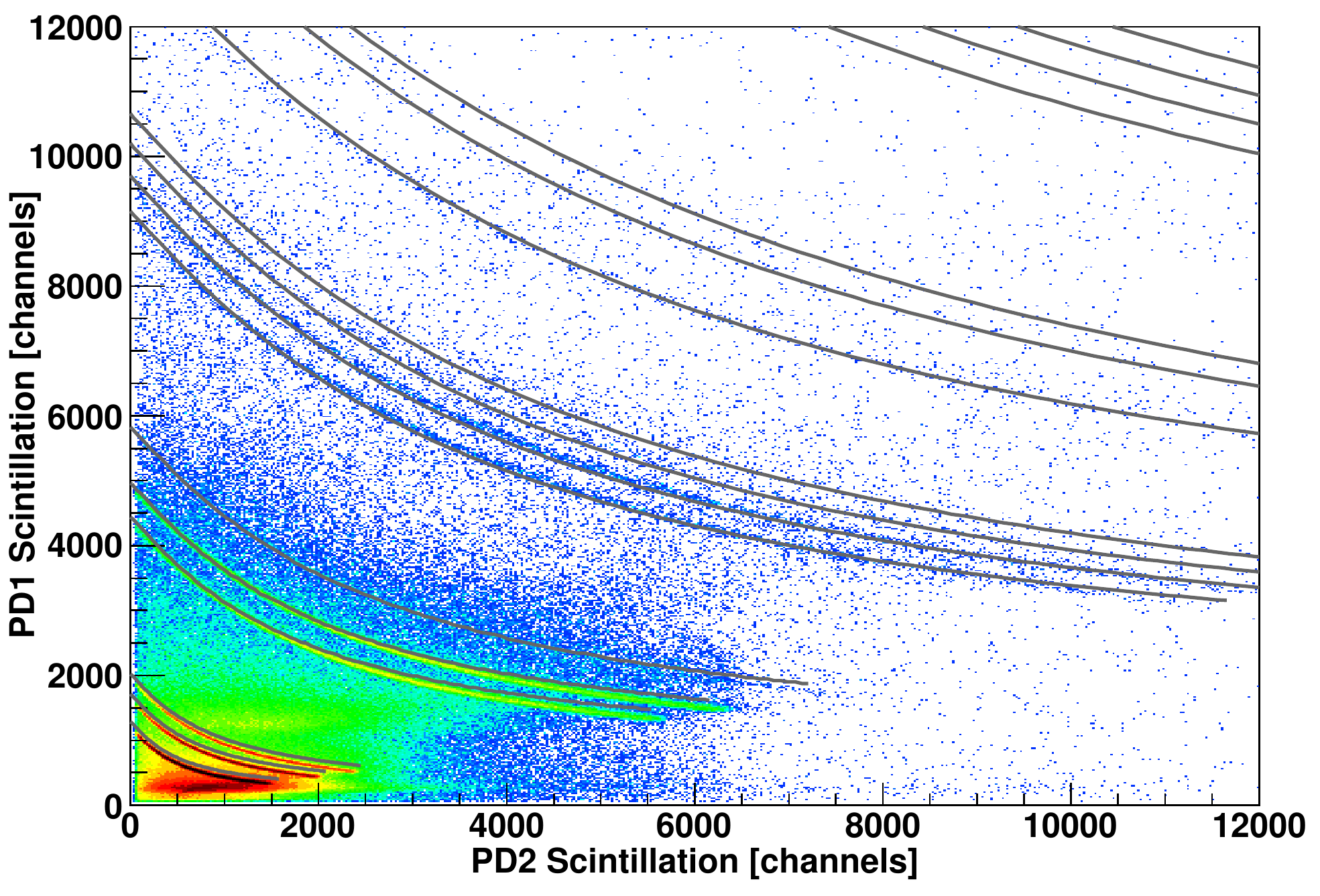} 

\caption{$\Delta$E-E identification map for scintillation signals from CsI1 vs
CsI2, with the  superimposed ID lines calculated using the ATIMA tables. The
sequence of lines is the same as in Fig. \ref{fig_raw01_atima}. The
$\gamma$-line and punch-through hits have been removed (cf. Fig.
\ref{fig_raw12}). }  

\label{fig_raw12_atima}
\end{figure}

Finally, Fig. \ref{fig_raw12_atima} shows the calculated ATIMA lines
superimposed on the $\Delta$E-E identification map for scintillation signals
from thin vs thick CsI(Tl) crystals. The slight overestimation of the $\Delta$E
component for $Z=1$ particles results most probably from the simplicity of the
Light-Energy conversion formula (\ref{eq_birks}) and (or) from the light output
non-uniformity of the crystals. The overall agreement is, nevertheless, quite
satisfactory and the calculated lines can be used not only to derive the energy
calibration parameters, but also for identification (see Fig.
\ref{fig_z_thick}), after small manual adjustments.

The energy calibration routine based on the ATIMA range-energy tables operated
on all three maps (Figs. \ref{fig_raw01_atima}-\ref{fig_raw12_atima})
simultaneously, allowing for a consistent determination of the calibration
parameters for all photodiodes and both crystals. Specifically, for the CsI(Tl)
crystals with 1500 ppm of Tl concentration, the quenching parameter $a_{2}$ was
found to be equal to 0.32, a value compatible with the one from \cite{horn} and
about 20\% larger than the average one quoted for the INDRA crystals
\cite{parlog}. The calibration of the SCT allowed also to estimate the
efficiency of the scintillation. The combination of a relatively high Tl doping,
high reflectance of the wrapping and large active area of the photodiode,
resulted in a relatively high efficiency for energy-light conversion. It was
found that only about 6 times more energy was needed to produce an electron-hole
pair in the photodiode through a scintillation process in the CsI(Tl) crystal
than in the direct ionization process in the photodiode. This fact is worth
noting, since for the ``good scintillators'' quoted in \cite{knoll} this factor
amounts to 15-20. 

Obviously, a drawback of the telescope method at high energies, is the high
level of the secondary reactions. In order to handle this problem, a more
sophisticated methods, neural networks and discriminant analysis are being
tested, but this goes beyond the scope of this article.   

\section{Summary} 

A new, low threshold, broad energy range, versatile array of triple telescopes,
KRATTA, has been constructed. The modules, equipped with digital electronics
chains, allowed for isotopic identification of light charged reaction products. 

Pulse shape analysis allowed for realistic decomposition of the complex SCT
pulse shapes into individual ionization and scintillation components and
eventually profit from the, otherwise harmful, nuclear counter effect. The
isotopic resolution obtained using a single readout channel was found to compete
very well with those obtained using the standard two channel readout. The
applied pulse shape analysis permitted also the identification of particles
stopped in the first photodiode and the reduction of the identification
threshold, due to the thickness of the first photodiode,  by a factor of three. 
Thanks to the pulse shape analysis, it was also possible to obtain the ballistic
deficit free amplitudes, which allowed for easy energy calibration and
identification based on the predictions of the range-energy tables.

The array has met the expectations, fulfilled the design requirements and
performed very well during the ASY-EOS experiment at GSI.

\section{Acknowledgments}

Work made possible through funding by Polish Ministry of Science and Higher
Education under grant No. DPN/N108/GSI/2009. 

We (S.K.) acknowledge the support by the Foundation for Polish Science - MPD
program, co-financed by the European Union within the European Regional
Development Fund.

Stimulating discussions, expertise and help of Giacomo Poggi as well as of
Marian P\^{a}rlog, Remi Bougault and Hector Alvarez Pol during the design and
test phase of the prototypes are gratefully acknowledged. 



\end{document}